\documentclass[%
 aip,
 pop,
 amsmath,amssymb,
preprint,%
nofootinbib,
]{revtex4-1}

\usepackage{graphicx}
\usepackage{mathrsfs}
\usepackage{amsmath}
\usepackage{natbib}
\usepackage{color}
\usepackage[normalem]{ulem}
\usepackage{hyperref}

\newcommand\bb[1]{\mbox{\boldmath{$#1$}}}
\newcommand\grad{\bb{\nabla}}

\newcommand\btimes{\,\bb{\times}\,}

\newcommand{\rmd}{{\rm d}}

\newcommand\Ev{\bb{E}}
\newcommand\Bv{\bb{B}}

\newcommand\Fextv{\bb{F}_{\rm ext}}
\newcommand\Jv{\bb{J}}
\newcommand\jv{\bb{j}}
\newcommand\uv{\bb{u}}

\newcommand\bv{\bb{b}}

\newcommand\xv{\bb{x}}
\newcommand\vv{\bb{v}}
\newcommand\wv{\bb{w}}
\newcommand\qv{\bb{q}}

\newcommand\bI{\bb{I}}

\newcommand\bQa{\bb{Q}_\alpha}

\newcommand\bQi{\bb{Q}_{\rm i}}

\newcommand\Omegaca{\Omega_{c\alpha}}
\newcommand\Omegace{\Omega_{c{\rm e}}}
\newcommand\Omegaci{\Omega_{c{\rm i}}}
\newcommand\ma{m_\alpha}
\newcommand\me{m_{\rm e}}
\newcommand\mi{m_{\rm i}}

\newcommand\nee{n_{\rm e}}
\newcommand\nii{n_{\rm i}}
\newcommand\uav{\bb{u}_\alpha}
\newcommand\uev{\bb{u}_{\rm e}}
\newcommand\uiv{\bb{u}_{\rm i}}
\newcommand\ppara{p_\|}
\newcommand\pperp{p_\perp}

\newcommand\pepara{p_{\|{\rm e}}}
\newcommand\peperp{p_{\perp{\rm e}}}

\newcommand\bPi{\bb{\Pi}}
\newcommand\bPia{\bb{\Pi}_\alpha}
\newcommand\bPie{\bb{\Pi}_{\rm e}}
\newcommand\bPii{\bb{\Pi}_{\rm i}}

\newcommand\vtha{v_{{\rm th},\alpha}}

\usepackage{scalerel,stackengine}
\stackMath
\newcommand\reallywidehat[1]{%
\savestack{\tmpbox}{\stretchto{%
  \scaleto{%
    \scalerel*[\widthof{\ensuremath{#1}}]{\kern-.6pt\bigwedge\kern-.6pt}%
    {\rule[-\textheight/2]{1ex}{\textheight}}%WIDTH-LIMITED BIG WEDGE
  }{\textheight}% 
}{0.5ex}}%
\stackon[1pt]{#1}{\tmpbox}%
}

\newcommand\reallywidetildeB[1]{
\stackon[-8pt]{#1}{\vstretch{1.5}{\hstretch{1.8}{\widetilde{\phantom{\;\;\;\;\;\;\;}}}}}
}

\newcommand\whFextv{\widehat{\bb{F}}_{\rm ext}}
\newcommand\whEv{\widehat{\bb{E}}}
\newcommand\whBv{\widehat{\bb{B}}}
\newcommand\whJv{\widehat{\bb{J}}}
\newcommand\wtvarrho{\widetilde{\varrho}}
\newcommand\whuav{\widehat{\bb{u}}_\alpha}
\newcommand\whuiv{\widehat{\bb{u}}_{\rm i}}
\newcommand\whuev{\widehat{\bb{u}}_{\rm e}}

\newcommand\wtbPii{\widetilde{\bb{\Pi}}_{\rm i}}
\newcommand\wtbPie{\widetilde{\bb{\Pi}}_{\rm e}}
\newcommand\wtbPia{\widetilde{\bb{\Pi}}_\alpha}
\newcommand\whbPii{\widehat{\bb{\Pi}}_{\rm i}}
\newcommand\whbPie{\widehat{\bb{\Pi}}_{\rm e}}

\newcommand\whPe{\widehat{P}_{\rm e}}
\newcommand\whbQi{\widehat{\bb{Q}}_{\rm i}}

\newcommand\Tnua{\bb{\cal T}_{nu}^{(\alpha)}}
\newcommand\Tnu{\bb{\cal T}_{nu}^{({\rm i})}}
\newcommand\Tnue{\bb{\cal T}_{nu}^{({\rm e})}}
\newcommand\TuaxB{\bb{\cal T}_{u\times B}^{(\alpha)}}
\newcommand\TuxB{\bb{\cal T}_{u\times B}^{({\rm i})}}
\newcommand\TuexB{\bb{\cal T}_{u\times B}^{({\rm e})}}
\newcommand\Tuaua{\bb{\cal T}_{uu}^{(\alpha)}}
\newcommand\Tuu{\bb{\cal T}_{uu}^{({\rm i})}}
\newcommand\Tueue{\bb{\cal T}_{uu}^{({\rm e})}}
\newcommand\whEu{\widehat{\cal E}_{u_{\rm i}}}
\newcommand\whEue{\widehat{\cal E}_{u_{\rm e}}}
\newcommand\whEua{\widehat{\cal E}_{u_\alpha}}
\newcommand\whEB{\widehat{\cal E}_B}
\newcommand\whEem{\widehat{\cal E}_\mathrm{em}}
\newcommand\whEPii{\widehat{\cal E}_{\Pi_{\rm i}}}
\newcommand\whEPie{\widehat{\cal E}_{\Pi_{\rm e}}}
\newcommand\whEPia{\widehat{\cal E}_{\Pi_\alpha}}
\newcommand\whEPe{\widehat{\cal E}_{P_{\rm e}}}
\newcommand\wtEPe{\widetilde{\cal E}_{P_{\rm e}}}
\newcommand\wtPe{\widetilde{P}_{\rm e}}
\newcommand\whqiv{\widehat{\bb{q}}_{\rm i}}
\newcommand\whqav{\widehat{\bb{q}}_\alpha}
\newcommand\bTPiinablau{\bb{\cal T}_{\Pi_{\rm i}\nabla u_{\rm i}}}

\newcommand\TPiinablau{{\cal T}_{\Pi\nabla u}^{({\rm i})}}
\newcommand\TPianablaua{{\cal T}_{\Pi\nabla u}^{(\alpha)}}
\newcommand\bTPienablaue{\bb{\cal T}_{\Pi\nabla u}^{({\rm e})}}

\newcommand\TPienablaue{{\cal T}_{\Pi\nabla u}^{({\rm e})}}
\newcommand\bTPiitimesB{\bb{\cal T}_{\Pi\times B}^{({\rm i})}}

\newcommand\bTPietimesB{\bb{\cal T}_{\Pi\times B}^{({\rm e})}}

\newcommand\TPenablaue{{\cal T}_{P\nabla u}^{({\rm e})}}

\newcommand\TJE{{\cal T}_{JE}}
\newcommand\TJEa{{\cal T}_{JE}^{(\alpha)}}
\newcommand\TJxB{\bb{\cal T}_{J\times B}}
\newcommand\TJJ{\bb{\cal T}_{JJ}}
\newcommand\TJuuJ{\bb{\cal T}_{[Ju]}}

\newcommand{\whcalSsg[1]}{\widehat{\cal S}_{\rm sg}^{(#1)}}
\newcommand{\whcalSsgtot}{\widehat{\cal S}_{\rm sg}}

\begin{document}

\title[Space-filtered equations for NHK models]{Space-filter techniques for quasi-neutral hybrid-kinetic models}

\author{S.~S.~Cerri}
\email[]{scerri@astro.princeton.edu}
%\homepage[]{https://www.astro.princeton.edu/~scerri/}
%\thanks{}
\affiliation{Department of Astrophysical Sciences, Princeton University, Princeton, NJ 08540, USA}

\author{E.~Camporeale}
\email[]{enrico.camporeale@noaa.gov}
%\thanks{}
\affiliation{CIRES, University of Colorado, Boulder, CO 80309, USA}
\affiliation{Centrum Wiskunde \& Informatica, 1098 XG Amsterdam, The Netherlands}

\date{\today}

\begin{abstract}
The space-filter approach has proved a fundamental tool in studying turbulence in neutral fluids, providing the ability to analyze scale-to-scale energy transfer in configuration space. It is well known that turbulence in plasma presents challenges different from neutral fluids, especially when the scale of interests include kinetic effects. The space-filter approach is still largely unexplored for kinetic plasma. Here we derive the space-filtered (or, equivalently ``coarse-grained'') equations in configuration space for a quasi-neutral hybrid-kinetic plasma model, in which ions are fully kinetic and electrons are a neutralizing fluid. 
Different models and closures for the electron fluid are considered, including finite electron-inertia effects and full electrons' pressure-tensor dynamics. 
Implications for the cascade of turbulent fluctuations in real space depending on different approximations are discussed.
\end{abstract}

\maketitle %\maketitle must follow title, authors, abstract and \pacs

\section{Introduction}\label{sec:Intro}

Turbulent plasmas can be found in a wide range of space, astrophysical and laboratory systems.
Understanding the properties of turbulent fluctuations and of their cascade -- which plasma processes allow the transfer of energy from the large (``injection'') scales to the small (``dissipative'') scales of a system, and how this energy is eventually converted into heat and non-thermal particles -- is a fundamental step in order to understand the evolution of such systems~\citep[e.g.,][]{QuataertGruzinovAPJ1999,SchekochihinCowleyPOP2006,KunzMNRAS2011,BrunoCarboneLRSP2013}.
In particular, turbulence in weakly collisional, magnetized plasmas is substantially different from turbulence in neutral fluids or in collisional plasmas, as it opens up the stage to a large variety of physical regimes~\citep[see, e.g.,][]{StawickiJGR2001,BoldyrevAPJL2005,SchekochihinAPJS2009,SahraouiPRL2010,camporeale2011dissipation,AlexandrovaSSRv2013,BoldyrevAPJ2013,ChenJPP2016,MatteiniMNRAS2017,KunzJPP2018,ChenBoldyrevAPJ2017,PassotSulem2019,VerscharenLRSP2019}.
This includes turbulent phase-space dynamics (``phase-space cascades'') and micro-instabilities~\citep[e.g.,][]{SchekochihinPPCF2008,camporeale2010implications,KunzPRL2014,HellingerAPJ2015,ServidioPRL2017,CerriAPJL2018,AdkinsSchekochihinJPP2018,Kawazura2018,PezziPOP2018,EyinkPRX2018,TeacaAPJ2019}.
A relevant aspect of plasma turbulence is the formation of localized (``coherent'') structures such as current sheets and magnetic structures, where both observations and simulations reveal an enhancement of kinetic features, from temperature anisotropy to particle energization and dissipation~\citep[e.g.,][]{ServidioNPG2011,GrecoPRE2012,LazarianPOP2012,OsmanPRL2012b,TenBargeAPJL2013,WuAPJL2013,ZhdankinAPJ2014,ChasapisAPJL2015,MatthaeusRSPTA2015,BanonNavarroPRL2016,FranciAIPC2016,ParasharMatthaeusAPJ2016,PerroneAPJ2016,PerroneAPJ2017,ValentiniNJP2016,WanPOP2016,GroseljAPJ2017,Califano2018,SorrisoValvoJPP2018,ArzamasskiyAPJ2019,WangAPJL2019}.
In this context, an increasing attention has been focused on the role of current sheets and of magnetic reconnection as possibly mediating the energy transfer in plasma turbulence, both in the framework of magneto-hydrodynamic (MHD) models and in the kinetic regime~\citep{CarbonePOF1990,haynes2014reconnection, BoldyrevLoureiroAPJ2017,LoureiroBoldyrevAPJ2017,MalletJPP2017,MalletMNRAS2017,CerriCalifanoNJP2017,FranciAPJL2017,CamporealePRL2018,ComissoAPJ2018,DongPRL2018,VechAPJL2018,PapiniAPJ2019}.
Therefore, it is of particular interest to develop a suitable theoretical framework that allows such investigation within kinetic and reduced-kinetic models that are widely adopted for kinetic-turbulence studies (see, e.g., \citet{CerriFrASS2019} and references therein).

The filtering approach has been intensively employed in the study of turbulence and scale-to-scale coupling in neutral fluids, at least since the seminal work of~\citet{germano1992}, and its subsequent use in Large Eddy Simulations (LES) both in the context of neutral fluids~\citep{MeneveauKatzARFM2000,carati2001} and MHD~\citep{agullo2001,chernyshov2006,aluie2010,YangPRE2016,YangPOF2017} (see, e.g., \citet{MieschSSRv2015} for a review on LES techniques in MHD). The approach is based on the following idea. First, a low-pass spatial filter is applied to all quantities of interest. Filtered quantities are then employed to construct equations for the conservation of density, momentum, and energy. Filtering is a linear operation (being essentially a convolution), meaning that, e.g., the filter applied to a product is not equal to the product of the filtered quantities. This implies that, in constructing energy equations, quadratic terms give rise to so-called sub-grid terms. The sub-grid terms play a crucial role in LES simulations, being the terms that lay below the resolved scale of the simulation and that are parameterized according to a given scheme~\citep[e.g.,][]{germano1991}. In this work, however, we follow the opposite philosophy, where sub-grid terms are actually well resolved by the computational grid. The rationale is that when the equation for total filtered energy is written in conservative form, the sub-grid terms represent a source/sink term that couples energy flux transfer at a given spatial scale. The interesting aspect of this approach, that has been exploited in \citet{CamporealePRL2018} to study correlations between energy dissipation and coherent structures, is that sub-grid terms are not defined in Fourier but in configuration space. One can then easily evaluate and study how the scale-to-scale energy transfer is related to other spatial-dependant quantities.
However, with a few notable exceptions~\citep[e.g.,][]{BanonNavarroPOP2014,YangPOP2017,YangMNRAS2019,CamporealePRL2018,EyinkPRX2018,kuzzay2018,BianAluiePRL2019}, the filtering approach has not received widespread attention in the plasma physics community.
For instance, in the context of 2.5D particle-in-cell (PIC) simulations plasma turbulence, this effort was undertaken by \citet{YangPOP2017}: by using the space-filtered techniques for their analysis, they found a qualitative coincidence between the spatial position of coherent structures and those sites with enhanced energy transfer.
Another important theoretical work is represented by \citet{EyinkPRX2018}, where the seminal idea to apply this formalism to the entire phase space was exploited for the full Vlasov-Maxwell-Landau equations. 
The present work is in between the original application to the MHD description (e.g., \citet{AluieNJP2017} and references therein) and the whole (six-dimensional) phase-space analysis of a full-kinetic description (e.g., \citet{EyinkPRX2018}).
In fact, here we consider a hybrid-kinetic model where the Vlasov equation for the ion distribution function is coupled to a neutralizing electron fluid (where different fluid models and a generalized Ohm's law are allowed). On the one hand, the hybrid approach will help to explicitly sort out and interpret the effects that are due to the different electron-ion dynamics, to different electron closures, and to their inertia, which would be of immediate interpretation otherwise. On the other hand, because of the richness of the terms arising in the resulting filtered equations already for this hybrid-kinetic case, for the sake of clearness of the analysis and discussion in the present work we will only consider the spatial part of the equations. 
Nevertheless, the configuration-space analysis will prove to be already able in highlighting extremely relevant differences and similarities between a fluid approach and a kinetic treatment of the turbulent cascade. For instance, in this work we provide a first formal explanation for the evidence that Hall-MHD and hybrid-kinetics with isothermal, massless electrons seem to provide similar results in terms of the turbulent cascade of magnetic-field fluctuations reported by \citet{PapiniAPJ2019}.
Therefore, an analysis of the whole coarse-grained phase space, as done, e.g., in \citet{EyinkPRX2018}, and of the mechanisms underlying the entire phase-space cascade of ion-entropy fluctuations in hybrid-kinetic models (see, e.g., \citet{CerriAPJL2018}) is out of the scope of the present study and is left for a future work.

The remainder of this paper is organized as follows.
In Section~\ref{sec:NHKmodel}, we present the equations of the general neutral hybrid-kinetic (NHK) model, including different closures on the electron fluid.
The energy equations for the (forced) NHK model are provided and discussed in Section~\ref{subsec:NHK_EnergyEqs}.
In Section~\ref{sec:NHK_EnEqs_SpaceFiltered}, we employ the so-called space-filtered techniques to the general NHK model equations.
The set of space-filtered energy equations are provided and discussed in Section~\ref{subsec:NHK_EnEqs_SpaceFiltered}.
The explicit set of equations for different versions of the hybrid-kinetic model that are often adopted in the literature is also provided in the Appendices \ref{app:sec:HVM} and \ref{app:sec:HK-resistive}. 
A straightforward generalization to the full-kinetic case is given in Appendix~\ref{app:sec:fullkinetic}.
Finally, in Section~\ref{sec:discussion-conclusion} we discuss the relevance of this theoretical framework for turbulent systems and collisionless plasma dynamics.

\section{The neutral hybrid-kinetic (NHK) model}\label{sec:NHKmodel}

The NHK model equations for a proton-electron plasma embedded in a magnetic field $\bb{B}$ can be written in the following form~\citep{TronciCamporealePOP2015}:
%%%%%%%%%%%%%%%%%%%%%%%%%%%%%%%%%%%%%%%
\begin{equation}\label{eq:NHK_Vlasov}
\frac{\partial\,f_{\rm i}}{\partial t}\, 
+\, \bb{v}\cdot\frac{\partial\,f_{\rm i}}{\partial \bb{x}}\, 
+\, \left[\frac{e}{m_{\rm i}}\Big(\bb{E} + \frac{\vv}{c}\times\bb{B}\Big) + \bb{F}_{\rm ext} \right] \cdot \frac{\partial\,f_{\rm i}}{\partial \bb{v}}\, =\, 0\,,
\end{equation}
\begin{equation}\label{eq:NHK_Ohm}
\Ev\,=\,
-\,\frac{\uev}{c}\times\Bv\,
-\,\frac{\grad\cdot\bPie}{en}\,
-\,\frac{\me}{e}\left[\frac{\partial\uev}{\partial t}+\big(\uev\cdot\grad\big)\uev\right]\,,
\end{equation}
\begin{equation}\label{eq:NHK_Maxwell}
\frac{\partial\,\bb{B}}{\partial t}\, =\, -\,c\,\grad\times\bb{E}\,,\quad
\bb{J}\,=\,\frac{c}{4\pi}\,\grad\times\bb{B}
\end{equation}
%%%%%%%%%%%%%%%%%%%%%%%%%%%%%%%%%%%%%%%
where $f_{\rm i}(\bb{x},\bb{v}, t)$ is the distribution function of the ions (protons) of mass $\mi$ at spatial position $\bb{x}$ and proton velocity $\bb{v}$, $e$ is the elementary charge, $c$ is the speed of light, and $\bb{F}_{\rm ext}$ is a external force per unit mass. In the generalized Ohm's law for the electric field $\bb{E}$, equation (\ref{eq:NHK_Ohm}), quasi-neutrality $\nee=\nii=n$ has been assumed, $\me$ is the electron mass, and the electron flow $\uev$ is related to the ion flow $\uiv$ and to the current density $\bb{J}$ by $\uev=\uiv-\Jv/en$. The above set of equations has to be closed by defining the thermal model of the electron fluid, e.g., by adopting a closure for the electron pressure tensor $\bPie$ or by providing evolution equations for its components. Notice that the form (\ref{eq:NHK_Ohm}) of the generalized Ohm's law is equivalent to the non-approximated version of its classical form (see Appendix~\ref{app:ohm} for details):
%%%%%%%%%%%%%%%%%%%%%%%%%%%%%%%%%%%%%%%
\begin{align}\label{eq:NHK_Ohm_2}
\Ev\,=\,&
\,-\frac{\uiv}{c}\times\Bv\,
+\,\frac{\Jv\times\Bv}{(1+\varepsilon_m)enc}\,
-\,\frac{\grad\cdot\big(\bPie-\varepsilon_m\bPii\big)}{(1+\varepsilon_m)en}\nonumber\\
\,&\,\nonumber\\
\,&\,+\frac{\varepsilon_m}{1+\varepsilon_m}\frac{\mi}{e^2n}\left[\frac{\partial\Jv}{\partial t}\,
+\,\grad\cdot\left(\Jv\uiv+\uiv\Jv-\frac{\Jv\Jv}{en}\right)\right]\,,
\end{align}
%%%%%%%%%%%%%%%%%%%%%%%%%%%%%%%%%%%%%%%
where $\varepsilon_m=\me/\mi\ll1$ is the (small) mass-ratio parameter. 

\subsection{The electron pressure equations and the fluid closure}

The NHK equations (\ref{eq:NHK_Vlasov})--(\ref{eq:NHK_Maxwell}) must be completed by one or more dynamic equations for the components of the electron pressure tensor, $\bPie$.
In the following, we consider the coupling with two models for the electron fluid:\\ 

\begin{itemize}
    \item[(a)] {\em isotropic, polytropic fluid}. This is the simplest case of a isotropic fluid, $\Pi_{{\rm e},ij}=P_{\rm e}\delta_{ij}$, with a polytropic closure~\citep[e.g.,][]{WinskeOmidiJGR1996,ValentiniJCP2007}
    %%%%%%%%%%%%%%%%%%%%%%%%%%%%%%%%%%%%%%%
    \begin{equation}\label{eq:NHK_Pe-polytropic}
     \frac{\rmd}{\rmd t}\left(\frac{P_{\rm e}}{n^\gamma}\right)\, =\, 0\,,
    \end{equation}
    %%%%%%%%%%%%%%%%%%%%%%%%%%%%%%%%%%%%%%%
    where $\gamma$ is the polytropic index (e.g., $\gamma=1$ for isothermal electrons), and the total (Lagrangian) derivative above uses the electron fluid velocity, i.e. $\rmd/\rmd t=\partial_t+\bb{u}_{\rm e}\cdot\nabla$. Therefore, using continuity equation, it rewrites as\footnote{Interestingly, by assuming a polytropic relation, $P_{\rm e}=C_\gamma n^\gamma$, Eq.~(\ref{eq:NHK_Pe-polytropic-2}) is actually the continuity equation for the electrons, as it can be rewritten as $\partial_t n + \grad\cdot(n\bb{u}_{\rm e})=0$, where quasi-neutrality has been assumed. The Vlasov equation also implies that $\partial_t n + \grad\cdot(n\bb{u}_{\rm i})=0$, and therefore  $\grad\cdot(n\bb{u}_{\rm i})=\grad\cdot(n\bb{u}_{\rm e})$ follows, i.e. $\grad\cdot\bb{J}=0$. This is consistently satisfied by $\bb{J}=\frac{c}{4\pi}\grad\times\bb{B}$.}
    %%%%%%%%%%%%%%%%%%%%%%%%%%%%%%%%%%%%%%%
    \begin{equation}\label{eq:NHK_Pe-polytropic-2}
     \frac{\partial\,P_{\rm e}}{\partial t}\,
     +\,\grad\cdot\big(P_{\rm e}\bb{u}_{\rm e}\big)\, =\, (1-\gamma)P_{\rm e}\big(\grad\cdot\bb{u}_{\rm e}\big)\,.
    \end{equation}
    %%%%%%%%%%%%%%%%%%%%%%%%%%%%%%%%%%%%%%%

    \item[(b)] {\em fully anisotropic, adiabatic fluid}. In this case, the full (agyrotropic) pressure tensor dynamics is retained and an adiabatic closure is adopted~\citep[e.g.,][]{CerriPOP2014b,DelSartoPPCF2017}:
    %%%%%%%%%%%%%%%%%%%%%%%%%%%%%%%%%%%%%%%
    \begin{align}\label{eq:NHK_PressTens-electrons}
     \frac{\partial\,\bPie}{\partial t}\, 
     +\, \grad\cdot\big(\bPie\uev\big)\,= 
     &-\, \Big\{\big(\bPie\cdot\grad\big)\uev\Big\}^{\rm sym}\nonumber\\
     &\,-\,\Omegace\,\big\{\bPie\times\bv\big\}^{\rm sym}\,,
    \end{align}
    %%%%%%%%%%%%%%%%%%%%%%%%%%%%%%%%%%%%%%%
    with the symmetrized terms, $\{\dots\}^{\rm sym}$, given by
    %%%%%%%%%%%%%%%%%%%%%%%%%%%%%%%%%%%%%%%
    \begin{equation}\label{eq:NHK_PressTens_sym1}
     \Big\{\big(\bPi\cdot\grad\big)\uv\Big\}_{ij}^{\rm sym}\,
     =\,\Pi_{ik}\partial_k u_j\, +\, \Pi_{jk}\partial_k u_i\,,
    \end{equation}
    %%%%%%%%%%%%%%%%%%%%%%%%%%%%%%%%%%%%%%%
    %%%%%%%%%%%%%%%%%%%%%%%%%%%%%%%%%%%%%%%
    \begin{equation}\label{eq:NHK_PressTens_sym2}
     \big\{\bPi\times\bv\big\}_{ij}^{\rm sym}\,
     =\,\epsilon_{ikl}\Pi_{jk}b_l\, +\, \epsilon_{jkl}\Pi_{ik}b_l\,,
    \end{equation}
    %%%%%%%%%%%%%%%%%%%%%%%%%%%%%%%%%%%%%%%
    where $\epsilon_{ijk}$ is the completely antisymmetric Levi-Civita symbol and $\bv=\Bv/|\Bv|$ is the unit vector along the magnetic field direction.
    Note that Eq.~(\ref{eq:NHK_PressTens-electrons}) reduces to the adiabatic version of Eq.~(\ref{eq:NHK_Pe-polytropic-2}), i.e. with $\gamma=5/3$, when an isotropic pressure tensor is considered, $\bPie=P_{\rm e}\bI$, and the equation is projected onto $\bI/3$. 
    It also provides the double-adiabatic limit when a gyrotropic pressure tensor is considered, $\bPie=\pepara\bv\bv+\peperp(\bI-\bv\bv)$, and the resulting equation is separately projected along the magnetic field direction and perpendicular to it, i.e. contracted with $\bv\bv$ and with $(\bI-\bv\bv)/2$, respectively~\citep[see, e.g.,][]{CGL1956,SnyderPOP1997}.
    In particular, note that a gyrotropic pressure tensor is also recovered in the limit of massless electrons, since equation (\ref{eq:NHK_PressTens-electrons}) requires that the condition $\big\{\bPie\times\bv\big\}^{\rm sym}=0$ is satisfied in such limit\footnote{The solution of $\big\{\bPie\times\bv\big\}^{\rm sym}=0$ is indeed the gyrotropic pressure tensor. This can be seen also in the context of a finite-Larmor-radius (FLR) expansion of the pressure tensor equation (see, e.g., \citet{CerriJPP2018} and references therein). In fact, the Larmor radius is proportional to the square root of the mass, $\rho_\alpha=\vtha/\Omegaca\propto\sqrt{m_\alpha}$, and therefore all the corrections to the gyrotropic pressure tensor vanish and only the secular evolution equations for the parallel and perpendicular components of the pressure, $\ppara$ and $\pperp$, are left.}.
    
\end{itemize} 

Let us now examine the energetics of the NHK model equations.

\subsection{Energy equations for the forced NHK system}\label{subsec:NHK_EnergyEqs}

The energy equations are derived using the moments of the Vlasov equation, i.e. from the corresponding fluid equations. Here, we interrupt the fluid hierarchy at the equation for the pressure tensor components:

%%%%%%%%%%%%%%%%%%%%%%%%%%%%%%%%%%%%%%%
\begin{equation}\label{eq:NHK_continuity}
\frac{\partial\,\varrho}{\partial t}\, +\, \grad\cdot\big(\varrho\uiv\big)\, =\, 0\,,
\end{equation}
%%%%%%%%%%%%%%%%%%%%%%%%%%%%%%%%%%%%%%%
%%%%%%%%%%%%%%%%%%%%%%%%%%%%%%%%%%%%%%%
\begin{equation}\label{eq:NHK_IonMomentum}
  \frac{\partial\,(\varrho\uiv)}{\partial t}\, +\, \grad\cdot\big(\varrho\uiv\uiv+\bPii\big)\, =\,
  \frac{e}{\mi}\varrho\left(\Ev+\frac{\uiv}{c}\times\Bv\right)\, +\, \varrho\Fextv\,,
\end{equation}
%%%%%%%%%%%%%%%%%%%%%%%%%%%%%%%%%%%%%%%
%%%%%%%%%%%%%%%%%%%%%%%%%%%%%%%%%%%%%%%
\begin{equation}\label{eq:NHK_PressTens}
  \frac{\partial\,\bPii}{\partial t}\, 
  +\, \grad\cdot\big(\bPii\uiv+\bQi\big)\, 
  +\, \Big\{\big(\bPii\cdot\grad\big)\uiv\Big\}^{\rm sym}=\,
  \Omegaci\,\big\{\bPii\times\bv\big\}^{\rm sym}\,.
\end{equation}
%%%%%%%%%%%%%%%%%%%%%%%%%%%%%%%%%%%%%%%
where $\varrho=m_{\rm i}n$ is the (ion) mass density (see Appendix \ref{app:FluidEqs_indexes} for explicit formulation with indexes and moments definitions). Note that as far as the global energy equations are concerned, the closure and/or the dynamic equation for the heat flux tensor is not relevant. In fact, it will enter the equations as a total divergence of a heat flux vector, $\grad\cdot\qv$ (see later), and for a ``closed'' system it will vanish when integrated over the whole system domain.
In this sense, the generalization of the following equations to the case where both ions and electrons are fully kinetic is straightforward (see Appendix~\ref{app:sec:fullkinetic}).

\subsubsection{Ion bulk (kinetic) energy}

By taking the scalar product between $\bb{u}_{\rm i}$ and the momentum equation, and using continuity equation, one finds the equation for the ion bulk energy:
%%%%%%%%%%%%%%%%%%%%%%%%%%%%%%%%%%%%%%%
\begin{equation}\label{eq:NHK_IonKinEn}
  \frac{\partial\,{\cal E}_{u_{\rm i}}}{\partial t}\, 
  +\, \grad\cdot\big({\cal E}_{u_{\rm i}}\uiv\big)\, 
  =\, -\uiv\cdot\big(\grad\cdot\bPii\big)\,
 +\,\varrho\uiv\cdot\left(\frac{e}{\mi}\Ev+\Fextv\right)\,,
\end{equation}
%%%%%%%%%%%%%%%%%%%%%%%%%%%%%%%%%%%%%%%
where ${\cal E}_{u_{\rm i}}=\frac{1}{2}\varrho u_{\rm i}^2$ is the ion bulk (or, kinetic) energy density.

\subsubsection{Ion internal (thermal) energy}

By taking the trace of Eq.~(\ref{eq:NHK_PressTens}), i.e. by contracting the indices (where a sum over repeated indices is understood), and multiplying by $1/2$, one obtains the equation for the ion internal energy:
\begin{equation}\label{eq:NHK_IonThermEn}
  \frac{\partial\,{\cal E}_{\Pi_{\rm i}}}{\partial t}\, 
  +\, \grad\cdot\big({\cal E}_{\Pi_{\rm i}}\uiv+\bb{q}_{\rm i}\big)\, 
  =\, -\bPii:\grad\uiv\,,
\end{equation}
%%%%%%%%%%%%%%%%%%%%%%%%%%%%%%%%%%%%%%%
where the ion internal (or, thermal) energy density is defined as ${\cal E}_{\Pi_{\rm i}}=\frac{1}{2}{\rm tr}[\bb{\Pi}_{\rm i}]$, and $\bb{q}_{\rm i}$ as the ion heat-flux vector defined by $q_{{\rm i},k}=\frac{1}{2}\sum_jQ_{{\rm i},jjk}$.

\subsubsection{Equivalent electron bulk (kinetic) energy}

By taking the scalar product between $en\uev$ and the generalized Ohm's law, Eq.~(\ref{eq:NHK_Ohm}), one obtains the equation for the electron bulk energy:
%%%%%%%%%%%%%%%%%%%%%%%%%%%%%%%%%%%%%%%
\begin{equation}\label{eq:NHK_ElectronKinEn}
  \frac{\partial\,{\cal E}_{u_{\rm e}}}{\partial t}\, 
  +\, \grad\cdot\big({\cal E}_{u_{\rm e}}\uev\big)\,
  =\, -\uev\cdot\big(\grad\cdot\bPie\big)\,
 -\,en\uev\cdot\Ev\,,
\end{equation}
%%%%%%%%%%%%%%%%%%%%%%%%%%%%%%%%%%%%%%%
where ${\cal E}_{u_{\rm e}}=\varepsilon_m\frac{1}{2}\varrho u_{\rm e}^2$ is the electron bulk (or, kinetic) energy density, and it vanishes in the limit of massless electrons, $\varepsilon_m\to0$.
In such limit, this equation becomes a statement of the balance between the work done on the electron fluid by the electric field and the electron pressure forces (in the electrons' reference frame\footnote{In fact, this is equivalent to use electric field in the frame of reference of the electron fluid, $\Ev'=\Ev+\uev\times\Bv/c$, since the $\uev\times\Bv$ term does not do any work on the electrons.}):
%%%%%%%%%%%%%%%%%%%%%%%%%%%%%%%%%%%%%%%
\begin{equation}\label{eq:NHK_ElectronKinEn-massless}
  en\uev\cdot\left(\Ev\,+\,\frac{\grad\cdot\bPie}{en}\right)\,=\,0\,,
\end{equation}
%%%%%%%%%%%%%%%%%%%%%%%%%%%%%%%%%%%%%%%
or, written in a way that may be useful later,
%%%%%%%%%%%%%%%%%%%%%%%%%%%%%%%%%%%%%%
\begin{equation}\label{eq:NHK_ElectronKinEn-massless-2}
  \grad\cdot\big(\bPie\cdot\uev\big)\,=\,
  \bPie:\grad\uev\,
  -\,en\uev\cdot\Ev\,.
\end{equation}
%%%%%%%%%%%%%%%%%%%%%%%%%%%%%%%%%%%%%%%
Note that this is true regardless of the assumptions on the pressure tensor of the electron fluid, as long as it is massless.

\subsubsection{Electron internal (thermal) energy}

We now derive the equation for the electron internal energy the two cases:\\

\begin{itemize}
    \item[(a)] {\em isotropic, polytropic fluid}. When the electron pressure tensor is diagonal and isotropic, i.e. $\Pi_{{\rm e},ij}=P_{\rm e}\delta_{ij}$, its associated internal energy is $\frac{1}{2}{\rm tr}[\bb{\Pi}_{\rm e}]=\frac{3}{2}P_{\rm e}$. Therefore, the corresponding energy equation is just the electron pressure equation, Eq.~(\ref{eq:NHK_Pe-polytropic-2}) , multiplied by $3/2$:  
    %%%%%%%%%%%%%%%%%%%%%%%%%%%%%%%%%%%%%%%
    \begin{equation}\label{eq:NHK_ElectronThermEn_polytropic}
     \frac{\partial\,{\cal E}_{P_{\rm e}}}{\partial t}\, 
     +\, \grad\cdot\big({\cal E}_{P_{\rm e}}\bb{u}_{\rm e}\big)\, 
     =\, (1-\gamma){\cal E}_{P_{\rm e}}\big(\grad\cdot\bb{u}_{\rm e}\big)\,,
    \end{equation}
    %%%%%%%%%%%%%%%%%%%%%%%%%%%%%%%%%%%%%%%
    where here ${\cal E}_{P_{\rm e}}=\frac{3}{2}P_{\rm e}$ is the electron internal energy density.
    
    \item[(b)] {\em fully anisotropic, adiabatic fluid}. In this case, the energy equation for the internal energy of the electron fluid is equivalent to the one of the ions, Eq.~(\ref{eq:NHK_IonThermEn}), with $\qv=0$:
    %%%%%%%%%%%%%%%%%%%%%%%%%%%%%%%%%%%%%%%
    \begin{equation}\label{eq:NHK_ElectronThermEn_fullPI}
     \frac{\partial\,{\cal E}_{\Pi_{\rm e}}}{\partial t}\, 
     +\, \grad\cdot\big({\cal E}_{\Pi_{\rm e}}\uev\big)\, 
     =\, -\bPie:\grad\uev\,,
    \end{equation}
    %%%%%%%%%%%%%%%%%%%%%%%%%%%%%%%%%%%%%%%
    with the electron internal energy density is now defined as ${\cal E}_{\Pi_{\rm e}}=\frac{1}{2}{\rm tr}[\bPie]$. Note that if $\bPie=P_{\rm e}\bI$, then Eq.~(\ref{eq:NHK_ElectronThermEn_fullPI}) correctly reduces to Eq.~(\ref{eq:NHK_ElectronThermEn_polytropic}) with $\gamma=5/3$.
\end{itemize}

\subsubsection{Magnetic energy}

By taking the scalar product between $\bb{B}$ and Faraday equation, and using the vector identity $\bb{B}\cdot(\grad\times\bb{E})=\grad\cdot(\bb{E}\times\bb{B})+\bb{E}\cdot(\grad\times\bb{B})$, one finds the equation for the ion bulk energy:
\begin{equation}\label{eq:NHK_MagneticEn-2}
  \frac{\partial\,{\cal E}_B}{\partial t}\, 
  +\, \grad\cdot\left(\frac{\bb{E}\times\bb{B}}{4\pi}c\right)\, 
  =\, -\bb{J}\cdot\bb{E}\,,
\end{equation}
%%%%%%%%%%%%%%%%%%%%%%%%%%%%%%%%%%%%%%%
%with ${\cal E}_B=B^2/8\pi$ being the magnetic energy density.
where ${\cal E}_B=B^2/8\pi$ is the magnetic energy density.
Note that in the hybrid approximation, there is no energy equation for the electric energy density, ${\cal E}_E=E^2/8\pi$, since the displacement current has been neglected\footnote{It is interesting to note that the $-\Jv\cdot\Ev$ term on the right-hand side of Eq.~(\ref{eq:NHK_MagneticEn-2}) comes from the approximated Amp\'ere's law, Eq.~(\ref{eq:NHK_Maxwell}). However, if the displacement current was retained, the equation for the magnetic energy density ${\cal E}_B=B^2/8\pi$ would read $\partial_t{\cal E}_B+\grad\cdot(\frac{c}{4\pi}\Ev\times\Bv)=-\frac{c}{4\pi}\Ev\cdot(\grad\times\Bv)$, and an additional equation for the electric energy density ${\cal E}_E=E^2/8\pi$ would be included, that can be written as $\partial_t{\cal E}_E=-\frac{c}{4\pi}\Ev\cdot(\grad\times\Bv)-\Jv\cdot\Ev$. Summing them together, we would get an equation for the total electromagnetic energy density, ${\cal E}_{\rm em}=(E^2+B^2)/8\pi$, that reads as (\ref{eq:NHK_MagneticEn-2}) with the substitution ${\cal E}_B\,\to\,{\cal E}_{\rm em}$ (see Appendix~\ref{app:sec:fullkinetic}).}.

\subsection{The total energy density equation and energy channels in the NHK model}

Let us gather all the previous energy equations here for convenience (and just rewriting few terms in a convenient way):
\begin{widetext}
%%%%%%%%%%%%%%%%%%%%%%%%%%%%%%%%%%%%%%%
\begin{align}
  \frac{\partial\,{\cal E}_{u_{\rm i}}}{\partial t}\, 
  +\, \grad\cdot\big({\cal E}_{u_{\rm i}}\bb{u}_{\rm i}+\bb{\Pi}_{\rm i}\cdot\bb{u}_{\rm i}\big)\, 
  =\, & \,\varrho\bb{u}_{\rm i}\cdot\bb{F}_{\rm ext}\,+\,\bb{\Pi}_{\rm i}:\grad\bb{u}_{\rm i}\,
  +\,en\bb{u}_{\rm i}\cdot\bb{E}\,,\label{eq:NHK_energy-summary-ionkin}\\
  %%%%
  \frac{\partial\,{\cal E}_{\Pi_{\rm i}}}{\partial t}\, 
  +\, \grad\cdot\big({\cal E}_{\Pi_{\rm i}}\bb{u}_{\rm i}+\bb{q}_{\rm i}\big)\, 
  =\, & \,-\bb{\Pi}_{\rm i}:\grad\bb{u}_{\rm i}\,,\label{eq:NHK_energy-summary-iontherm}\\
  %%%%
  \frac{\partial\,{\cal E}_{u_{\rm e}}}{\partial t}\, 
  +\, \grad\cdot\big({\cal E}_{u_{\rm e}}\uev+\bPie\cdot\uev\big)\, 
  =\, & \,\bPie:\grad\uev\,
  -\,en\uev\cdot\Ev\,,\label{eq:NHK_energy-summary-eleckin}\\
  %%%%
  \frac{\partial\,{\cal E}_{\Pi_{\rm e}}}{\partial t}\, 
  +\, \grad\cdot\big({\cal E}_{\Pi_{\rm e}}\bb{u}_{\rm e}\big)\, 
  =\,& \left\{\begin{array}{l}
       (1-\gamma){\cal E}_{P_{\rm e}}\big(\grad\cdot\bb{u}_{\rm e}\big)\\
       \,\\
      -\bPie:\grad\uev
  \end{array}\right.\begin{array}{c}
        \qquad\textrm{(a)}\\
        \, \\
        \qquad\textrm{(b)}
  \end{array}\,,\label{eq:NHK_energy-summary-electherm}\\
  %%%%
  \frac{\partial\,{\cal E}_B}{\partial t}\, 
  +\, \grad\cdot\left(\frac{\bb{E}\times\bb{B}}{4\pi}c\right)\, 
  =\, & \,-\,en\bb{u}_{\rm i}\cdot\bb{E}\,+\,en\bb{u}_{\rm e}\cdot\bb{E}\label{eq:NHK_energy-summary-magn}\,,
\end{align}
%%%%%%%%%%%%%%%%%%%%%%%%%%%%%%%%%%%%%%%
\end{widetext}
where we remind the reader the following definitions:
%%%%%%%%%%%%%%%%%%%%%%%%%%%%%%%%%%%%%%%
\begin{equation}\label{eq:HVM_En-defs-all}
  {\cal E}_{u_\alpha}\,=\,\frac{1}{2}\ma n u_\alpha^2\,,\quad
  {\cal E}_{\Pi_\alpha}=\,\frac{1}{2}{\rm tr}[\bPia]\,,\quad
  {\cal E}_B=\,\frac{B^2}{8\pi}\,.
\end{equation}
%%%%%%%%%%%%%%%%%%%%%%%%%%%%%%%%%%%%%%%
Note that equations (\ref{eq:NHK_energy-summary-ionkin})--(\ref{eq:NHK_energy-summary-magn}) are well-known fundamental energy equations describing the energy channels of a kinetic plasma, and essentially the same set of equations can be derived for the full-Vlasov case~\citep{KrallTrivelpieceBOOK1973,YangPRE2017,YangPOP2017,YangMNRAS2019}.
By summing up equations (\ref{eq:NHK_energy-summary-ionkin})--(\ref{eq:NHK_energy-summary-magn}), one obtains the equation for the total energy density, ${\cal E}={\cal E}_{u_{\rm i}}+{\cal E}_{\Pi_{\rm i}}+{\cal E}_{u_{\rm e}}+{\cal E}_{\Pi_{\rm e}}+{\cal E}_B$:
%%%%%%%%%%%%%%%%%%%%%%%%%%%%%%%%%%%%%%%
\begin{equation}\label{eq:NHK_TotalEn-1}
  \frac{\partial\,{\cal E}}{\partial t}\, 
  +\, \grad\cdot\bb{\Phi}_{\cal E}\, 
  =\, \varrho\uiv\cdot\Fextv\,
  +\,{\cal I}_{\rm e}\,,
\end{equation}
%%%%%%%%%%%%%%%%%%%%%%%%%%%%%%%%%%%%%%%
where we have defined the energy density flux, $\bb{\Phi}_{\cal E}$, as
%%%%%%%%%%%%%%%%%%%%%%%%%%%%%%%%%%%%%%%
\begin{equation}
  \bb{\Phi}_{\cal E}\,=\,
  \big({\cal E}_{u_{\rm i}}+{\cal E}_{\Pi_{\rm i}}\big)\uiv\,
  +\,\bPii\cdot\uiv\,
  +\,\bb{q}_{\rm i} \,+\big({\cal E}_{u_{\rm e}}+{\cal E}_{\Pi_{\rm e}}\big)\uev\,
  +\,\bPie\cdot\uev\,
  +\,\frac{\bb{E}\times\bb{B}}{4\pi}c\,.\label{eq:NHK_TotalEnFlux-1}
\end{equation}
%%%%%%%%%%%%%%%%%%%%%%%%%%%%%%%%%%%%%%%
The additional ${\cal I}_{\rm e}$ term comes from the closure adopted for the electron fluid, and it is zero for an adiabatic fluid:
%%%%%%%%%%%%%%%%%%%%%%%%%%%%%%%%%%%%%%%
\begin{equation}\label{eq:NHK_Ie-term}
  {\cal I}_{\rm e}\, 
  =\, 
  \left\{\begin{array}{l}
    \left(\frac{5}{3}-\gamma\right){\cal E}_{P_{\rm e}}\big(\grad\cdot\uev\big)    \\
    \,  \\  
    0    
  \end{array}\right.\begin{array}{c}
        \qquad\textrm{(a)}\\
        \, \\
        \qquad\textrm{(b)}
  \end{array}\,.
\end{equation}
%%%%%%%%%%%%%%%%%%%%%%%%%%%%%%%%%%%%%%%
From the above equations is clear that when an isothermal closure is adopted for the electron fluid, there cannot be an exact conservation of energy even when no external injection is considered, $\Fextv=0$. In fact, artificial energy loss or gain has to be included in the model in order to keep the electrons isothermal when compressibility effects are present (i.e., when $\grad\cdot\uev=\grad\cdot\uiv+(\Jv/en)\cdot\grad\ln(n)\neq0$).
This is indeed a well-known feature in the astrophysics community, and it is often attributed to instantaneous radiation cooling and heating (e.g., turbulent heating) that on average maintain the isothermal state~\citep{Nordlund_1999,Vazquez-Semadeni_1999,Vazquez-Semadeni_2000}.

Finally, it is interesting to note the coupling between the different energy densities in equations (\ref{eq:NHK_energy-summary-ionkin})--(\ref{eq:NHK_energy-summary-magn}), i.e. the so-called ``energy channels''. 
The thermal energy only couples to the kinetic energy via the pressure-strain term, $\bPia:\grad\uav$ (with some modifications when explicit closure relations are adopted on the pressure), whereas the magnetic energy is only coupled to the kinetic energy via the electric-field work term, $e_\alpha n_\alpha\uav\cdot\Ev$. 
Moreover, as it is expected in a collisionless system, the ions' and electrons' energy channels never couple to each other directly, but only through electromagnetic fields.
This is more evident if we take a spatial average over the entire spatial domain under consideration, denoted by $\langle\dots\rangle$, and we assume that the fluxes are such that $\langle\grad\cdot(\dots)\rangle=0$, as, e.g., for periodic or insulating boundary conditions~\citep{YangPRE2017,YangPOP2017,YangMNRAS2019}:
%%%%%%%%%%%%%%%%%%%%%%%%%%%%%%%%%%%%%%%
\begin{align}
  \frac{\partial\,\langle{\cal E}_{u_{\rm i}}\rangle}{\partial t}\, 
  =\, & \,\langle\varrho\uiv\cdot\Fextv\rangle\,
  +\,\langle\bPii:\grad\uiv\rangle\,
  +\,\langle\bb{j}_\mathrm{i}\cdot\Ev\rangle\,,\label{eq:NHK_energy-summary-ionkin-avg}\\
  %%%%
  \frac{\partial\,\langle{\cal E}_{\Pi_{\rm i}}\rangle}{\partial t}\, 
  =\, & \,-\langle\bPii:\grad\uiv\rangle\,,\label{eq:NHK_energy-summary-iontherm-avg}\\
  %%%%
  \frac{\partial\,\langle{\cal E}_{u_{\rm e}}\rangle}{\partial t}\, 
  =\, & \,\langle\bPie:\grad\uev\rangle\,
  +\,\langle\bb{j}_\mathrm{e}\cdot\Ev\rangle\,,\label{eq:NHK_energy-summary-eleckin-avg}\\
  %%%%
  \frac{\partial\,\langle{\cal E}_{\Pi_{\rm e}}\rangle}{\partial t}\, 
  =\,& \left\{\begin{array}{l}
       (1-\gamma)\langle{\cal E}_{P_{\rm e}}\big(\grad\cdot\bb{u}_{\rm e}\big)\rangle\\
       \,\\
      -\langle\bPie:\grad\uev\rangle
  \end{array}\right.\begin{array}{c}
        \qquad\textrm{(a)}\\
        \, \\
        \qquad\textrm{(b)}
  \end{array}\,,\label{eq:NHK_energy-summary-electherm-avg}\\
  %%%%
  \frac{\partial\,\langle{\cal E}_B\rangle}{\partial t}\, 
  =\, & \,-\langle\bb{j}_\mathrm{i}\cdot\Ev\rangle\,-\,\langle\bb{j}_\mathrm{e}\cdot\Ev\rangle\label{eq:NHK_energy-summary-magn-avg}\,,
\end{align}
%%%%%%%%%%%%%%%%%%%%%%%%%%%%%%%%%%%%%%%
where we have introduced the species' current density, $\bb{j}_\alpha\equiv e_\alpha n\uav$, for shortness.
Therefore, kinetic energy acts as the only mediator in the transfer between the field and the thermal energies, and conversion from electromagnetic energy to internal energy of the plasma has to necessarily go through the generation of bulk flows~\citep{YangPRE2017,YangPOP2017,YangMNRAS2019}.  
At the same time, in a collisionless system, the energy transfer from one species to another can only occur through field energy. Thus, in steady state conditions, electromagnetic fields are the mediators that determine the partition between the species' energy cascades.
However, the above equations are volume averaged, and so they do not contain information about the spatial regions to which the energy transfer is associated or about scale-by-scale energy transfer and cross-scale interactions: in order to investigate these properties, special techniques need to be applied~\citep[see, e.g.,][]{CamporealePRL2018}. 
The space-filtered equations for the general NHK model (\ref{eq:NHK_Vlasov})--(\ref{eq:NHK_Maxwell}), with both type of closures, (\ref{eq:NHK_Pe-polytropic}) or (\ref{eq:NHK_PressTens-electrons}), are derived in the next Section, while the equations belonging to specific hybrid-kinetic (HK) models are provided in Appendix \ref{app:sec:HVM} and \ref{app:sec:HK-resistive}.
The full-kinetic (FK) case is also provided in Appendix~\ref{app:sec:fullkinetic}.

\section{The space-filtered approach to the general NHK model}\label{sec:NHK_EnEqs_SpaceFiltered}

Following the procedure described in \citet{CamporealePRL2018}, we now derive the corresponding filtered equations for the energy densities in the NHK system. 

\subsection{Filter definitions and properties}\label{subsec:filter-def}

Let us consider a vector field, $\bb{V}(\xv,t)$. The corresponding space-filtered field $\widetilde{\bb{V}}(\xv,t)$ is defined as
%%%%%%%%%%%%%%%%%%%%%%%%%%%%%%%%%%%%%%%
\begin{equation}\label{eq:filter-def-1}
 \widetilde{\bb{V}}(\xv,t)\,=\,\int_{\Omega}G(\xv-\bb{\xi})\bb{V}(\bb{\xi},t)\rmd^3\bb{\xi}\,=\,\bb{V}(\xv,t)\star G(\xv)\,,
\end{equation}
%%%%%%%%%%%%%%%%%%%%%%%%%%%%%%%%%%%%%%%
where $\Omega$ is the entire spatial domain, $G$ is the filter function (e.g., Gaussian, top-hat, Fourier, etc), and $\star$ is the convolution operator defined by Eq.~(\ref{eq:filter-def-1}). 
We also assume that the filtering operation commutes with time and space differentiation:
%%%%%%%%%%%%%%%%%%%%%%%%%%%%%%%%%%%%%%%
\begin{align}\label{eq:filter-def-2}
 \widetilde{\frac{\rmd\bb{V}}{\rmd t}}\,=\, & \,\frac{\rmd\widetilde{\bb{V}}}{\rmd t}\,,\\
 \widetilde{\grad\cdot\bb{V}}\,=\, & \,\grad\cdot\widetilde{\bb{V}}\,.
\end{align}
%%%%%%%%%%%%%%%%%%%%%%%%%%%%%%%%%%%%%%%
Finally, we introduce the so-called Favre filter~\citep{FavrePOF1983}, defined as
%%%%%%%%%%%%%%%%%%%%%%%%%%%%%%%%%%%%%%%
\begin{equation}\label{eq:filter-def-3}
 \widehat{\bb{V}}\,=\, \frac{\widetilde{\varrho\bb{V}}}{\widetilde{\varrho}}\,.
\end{equation}
%%%%%%%%%%%%%%%%%%%%%%%%%%%%%%%%%%%%%%%
This type of density-weighted filtering operation was originally introduced in hydrodynamic turbulence in order to avoid possible contamination due to the viscous dynamics in the large-scales quantities, and thus allowing for the existence of an inertial-range cascade (``inviscid criterion''~\citep{AluiePhyD2013,ZhaoAluiePRF2018}). Although in the present case we are analyzing formally collisionless equations, i.e., without any effective viscosity or resistivity, in the following we will make use of this filtering procedure. A practical reason to follow such approach is that the underlying idea of this work is to later apply the equations obtained with this formalism to actual numerical simulations, which will necessarily present some artificial dissipation mechanisms (e.g., numerical smoothing, hyper-dissipation operators, or even actual resistivity in the Ohm's law -- see, e.g., Appendix~\ref{app:sec:HK-resistive}). Therefore, by using the Favre-filtering approach, we should ensure the applicability to the analysis of hybrid-kinetic numerical simulations.

\subsection{Space-filtered energy equations for the forced NHK system}\label{subsec:NHK_EnEqs_SpaceFiltered}

We start by filtering the single energy equations, i.e., by considering the energy conservation associated to those scales larger than a given {\em filtering scale} $\ell$. In this process, we will also define certain {\em sub-grid terms} arising from non-linear terms in the equation, such as, for instance,
%%%%%%%%%%%%%%%%%%%%%%%%%%%%%%%%%%%%%%%
\begin{equation}\label{eq:sub-grid-def_example}
\widetilde{\,\varrho\uiv\uiv}\,
=\,\widetilde{\varrho}\,\widehat{\uiv\uiv}\,
=\,\widetilde{\varrho}\,\whuiv\whuiv+\Tuu\,,
\end{equation}
%%%%%%%%%%%%%%%%%%%%%%%%%%%%%%%%%%%%%%%
i.e., the sub-grid $\Tuu$ term is determined by the difference of the non-linear terms
%%%%%%%%%%%%%%%%%%%%%%%%%%%%%%%%%%%%%%%
\begin{equation}\label{eq:NHK_Tuu-def-1}
\Tuu\,\equiv\,
\widetilde{\varrho}\big(\widehat{\uiv\uiv}-\whuiv\whuiv\big)\,,
\end{equation}
%%%%%%%%%%%%%%%%%%%%%%%%%%%%%%%%%%%%%%%
and represents term that is associated to the ion-flow non-linearity at scales $<\ell$.

\subsubsection{Filtered ion bulk (kinetic) energy equation}

By filtering the ion momentum equation (\ref{eq:NHK_IonMomentum}) and using the above definitions, one gets the filtered momentum equation in the following form:
%\begin{widetext}
%%%%%%%%%%%%%%%%%%%%%%%%%%%%%%%%%%%%%%%
\begin{align}\label{eq:NHK_IonMomentum-filtered}
  \frac{\partial\,(\wtvarrho\whuiv)}{\partial t}+
  \grad\cdot\big(\wtvarrho\whuiv\whuiv\big)=&
  \,-\grad\cdot\big(\Tuu+\wtbPii\big)\,
  +\,\wtvarrho\whFextv\nonumber\\
  &+\frac{e}{m_{\rm i}}\wtvarrho\left(\whEv+\frac{\whuiv}{c}\times\whBv+\TuxB\right)\,,
\end{align}
%%%%%%%%%%%%%%%%%%%%%%%%%%%%%%%%%%%%%%%
%\end{widetext}
where the sub-grid term related to the ion-flow non-linearity, $\Tuu$, is defined in (\ref{eq:NHK_Tuu-def-1}),
and we have introduced the sub-grid term associated with the $\uiv\times\Bv$ non-linearity,
%%%%%%%%%%%%%%%%%%%%%%%%%%%%%%%%%%%%%%%
\begin{equation}\label{eq:NHK_TuxB-def-1}
  \TuxB\,=\,\frac{1}{c}\Big(
  \reallywidehat{\uiv\times\Bv}-\whuiv\times\whBv\Big)\,.
\end{equation}
%%%%%%%%%%%%%%%%%%%%%%%%%%%%%%%%%%%%%%%
By taking the scalar product of (\ref{eq:NHK_IonMomentum-filtered}) with $\whuiv$, the filtered equation for the ion bulk energy density follows\citep{YangPOP2017}:
%%%%%%%%%%%%%%%%%%%%%%%%%%%%%%%%%%%%%%%
\begin{align}\label{eq:NHK_IonKinEn-filtered}
  \frac{\partial\,\whEu}{\partial t} +\, 
  \grad\cdot\big(\whEu\whuiv\big)=\,
  &\,\wtvarrho\,\whuiv\cdot\Big[\whFextv+\frac{e}{\mi}\big(\whEv+\TuxB\big)\Big]\nonumber\\
  &\,-\whuiv\cdot\Big[\grad\cdot\big(\wtbPii+\Tuu\big)\Big]\,,
\end{align}
%%%%%%%%%%%%%%%%%%%%%%%%%%%%%%%%%%%%%%%
where now the filtered ion bulk (kinetic) energy density is defined as $\whEu=\frac{1}{2}\wtvarrho|\whuiv|^2$.

Note that the sub-grid terms above have an immediate physical interpretation as, e.g., ``turbulent'' fields or stresses.
If $\ell$ is a characteristic scale of the filter defined in (\ref{eq:filter-def-1}), then the first sub-gird term, $\Tuu$, is indeed the Reynolds stress associated with the ion-flow fluctuations at scales $<\ell$ (i.e., what contributes to the transport of ion-momentum at scales $\ell$, $\wtvarrho\,\whuiv$, due to the advection of
sub-scale ion-momentum by sub-scale ion-flow fluctuations~\citep{AluieNJP2017}, in analogy to its original definitions within mean-field MHD~\citep{Moffatt_BOOK1978,KrauseRaedler_BOOK1980}).
Analogously, $\TuxB$ is (minus) the ``MHD contribution'' to the ``turbulent'' electric field, $\bb{\epsilon}_\mathrm{MHD}^*\equiv-\TuxB$, i.e., the electric field associated to the $\uiv\times\Bv$ fluctuations at scales $<\ell$ (see Section~\ref{subsec:NHK_E-field-filtered}).
These terms, i.e., the ``sub-scale electromotive foce'' $\bb{\epsilon}_\mathrm{MHD}^*$ and the ``sub-scale Reynolds stress'' $\Tuu$ associated to the ion-flow fluctuations, are indeed analogous to those present in the coarse-grained MHD equations (see, e.g., \citet{AluieNJP2017} and references therein).
Thus, another useful way to rewrite the equation above is
\begin{widetext}
%%%%%%%%%%%%%%%%%%%%%%%%%%%%%%%%%%%%%%%
\begin{equation}\label{eq:NHK_IonKinEn-filtered-1bis}
  \frac{\partial\,\whEu}{\partial t}\, +\, 
  \grad\cdot\Big[\whEu\whuiv+\big(\wtbPii+\Tuu\big)\cdot\whuiv\Big]\, =\, 
  \wtvarrho\,\whuiv\cdot\whFextv\,
  +\,\widehat{\bb{j}}_\mathrm{i}\cdot\whEv\,
  +\,\wtbPii:\grad\whuiv\, -\,\widehat{\bb{j}}_\mathrm{i}\cdot\bb{\epsilon}_\mathrm{MHD}^*\,
 +\,\Tuu:\grad\whuiv\,,
\end{equation}
%%%%%%%%%%%%%%%%%%%%%%%%%%%%%%%%%%%%%%%
\end{widetext}
where we have introduced the Favre-filtered ion current density, $\widehat{\bb{j}}_\mathrm{i}\equiv\frac{e}{\mi}\wtvarrho\,\whuiv=e\widetilde{n}\whuiv$.
Therefore, when considering a characteristic scale $\ell$ for the filters, the energy transfer of ions' kinetic energy through that scale is mediated by a combination of two effects. 
The first is represented by the interaction between the ion current density at scales $\geq\ell$, $\widehat{\bb{j}}_{\mathrm{i},\ell}$, and the MHD contribution of the ``turbulent'' electric-field fluctuations at scales $<\ell$, $\bb{\epsilon}_\mathrm{MHD}^*$.
The second effect relies on the interaction of the strain tensor of the ion flow at scales $\geq\ell$, $\widehat{\bb{\Sigma}}_{\mathrm{i},\ell}\equiv\grad\whuiv$, with the ``turbulent'' ion-flow Reynolds stress at scales $<\ell$, $\bb{\mathcal{T}}_{uu,\ell}^{(\mathrm{i})}$.
That is, the net ion-kinetic energy density flux through a scale $\ell$, $\mathcal{T}_{\mathrm{kin},\ell}^{(\mathrm{i})}$, is given by the combination of the above two terms: $\mathcal{T}_{\mathrm{kin},\ell}^{(\mathrm{i})}=\bb{\mathcal{T}}_{uu,\ell}^{(\mathrm{i})}:\widehat{\bb{\Sigma}}_{\mathrm{i},\ell}-\widehat{\bb{j}}_{\mathrm{i},\ell}\cdot\bb{\epsilon}_{\mathrm{MHD},\ell}^*$.
The strain-Reynolds stresses interaction as a mechanism for scale-to-scale energy transfer is indeed analogous to the one arising in the coarse-grained MHD equations~\citep{AluieNJP2017,BianAluiePRL2019}. The term $\widehat{\bb{j}}_{\mathrm{i},\ell}\cdot\bb{\epsilon}_{\mathrm{MHD},\ell}^*$ represents the ionic contribution to the so-called ``turbulent Ohmic dissipation'' due to the MHD sub-grid electromotive force associated to the $\uiv\btimes\Bv$ fluctuations, {\em viz.} $\widehat{\bb{J}}_\ell\cdot\bb{\epsilon}_{\mathrm{MHD},\ell}^*$, that would also be present in the MHD description~\citep{AluieNJP2017,BianAluiePRL2019}. As we will discuss in the following Sections, however, in the hybrid-kinetic case there are additional (i.e., non-MHD) contributions to the 
``sub-grid electric field'' $\bb{\epsilon}_\ell^*$, as well as sub-grid terms arising from a two-fluid-like description (i.e., ions and electrons, as opposed to the single-fluid MHD treatment). This in turn provides additional scale-to-scale energy transfer mechanisms through, e.g., turbulent Ohmic dissipation and strain-Reynolds stresses interaction of the electron fluid (see Section~\ref{subsec:NHKtotalenergy_filtered}).
The term $\wtvarrho\whuiv\cdot\whFextv$ is the rate of ion-kinetic energy density injected into the system by the external forcing at scales $\geq\ell$.
If the forcing acts only at scales $\ell_F$ and $\Fextv$ itself is a constant, then this term is a constant, $\varepsilon_{F,0}$, for all the scales $\ell<\ell_F$. However, we caution that in a more general case, if $\Fextv$ varies at scales $\ell_F$ and the density is not a constant, then it has been shown that $\varepsilon_F$ is only approximately constant: while decaying at scales $\ell<\ell_F$, it is not exactly zero (see, e.g., \citet{AluiePhyD2013}).
The other terms on the first line of the right-hand side of (\ref{eq:NHK_IonKinEn-filtered-1bis}) instead represent the energy density flux between the different energy channels at scales $\geq\ell$.
In particular, the ion-kinetic energy is connected to the magnetic-field energy channel through the interaction between the ion-current density and the electric field at scales $\geq\ell$, $\widehat{\bb{j}}_{\mathrm{i},\ell}\cdot\whEv_\ell$, and to the ion-thermal energy through the pressure-strain interaction at scales $\geq\ell$, $\widetilde{\bb{\Pi}}_{\mathrm{i},\ell}:\widehat{\bb{\Sigma}}_{\mathrm{i},\ell}$ (see later). 
Also, note that for a deeper investigation of the pressure-strain interaction at scales $\geq\ell$, one can further decompose the strain tensor into different contributions, i.e., isotropic compression and volume-preserving deformations and rotations, $\bb{\Sigma}=-\mathcal{C}\bb{I}+\bb{\mathcal{D}}+\bb{\mathcal{W}}$~\citep[see, e.g.][]{YangPOP2017,DelSartoPRE2016,DelSartoPegoraroMNRAS2018}).

\subsubsection{Filtered ion internal (thermal) energy equation}

We now apply a Favre filter on the ion pressure tensor equation, so we obtain
%%%%%%%%%%%%%%%%%%%%%%%%%%%%%%%%%%%%%%%
\begin{align}
  \frac{\partial\,\whbPii}{\partial t} 
  + \grad\cdot\big(\widehat{\bPii\uiv}+\whbQi\big)=
  & \,-\Big\{\big(\wtbPii\cdot\grad\big)\whuiv+\bTPiinablau\Big\}^{\rm sym}\nonumber\\
  & \,+\frac{e}{\mi c}\Big\{\wtbPii\times\whBv+\bTPiitimesB\Big\}^{\rm sym}\,,\label{eq:NHK_PressTens-filtered}
\end{align}
%%%%%%%%%%%%%%%%%%%%%%%%%%%%%%%%%%%%%%%
where we have introduced the sub-grid tensors associated to the pressure-strain and pressure-magnetic ions' non-linearities\footnote{Following \citet{DelSartoPegoraroMNRAS2018}, one could construct a magnetic matrix $\boldsymbol{\cal B}$ related to the local magnetic field by duality, ${\cal B}_{ij}=\epsilon_{ijk}B_k$, and write this non-linear interaction as the commutator $[\boldsymbol{\cal B},\bPi]$. A similar procedure could be used to describe the non-linear interaction between the pressure tensor $\bPi$ and the strain tensor $\boldsymbol{\Sigma}$, where $\Sigma_{ij}=\partial_i u_j$. See also \citet{DelSartoPRE2016} or \citet{DelSartoPegoraroMNRAS2018} for further decomposition of $\boldsymbol{\Sigma}$ into different contributions (i.e., isotropic compression, volume-preserving deformations and rotations\citep{DelSartoPRE2016,YangPRE2017,YangPOP2017,DelSartoPegoraroMNRAS2018,YangMNRAS2019}) in terms of commutators and anti-commutators.}:
%%%%%%%%%%%%%%%%%%%%%%%%%%%%%%%%%%%%%%%
\begin{equation}\label{eq:NHK_bTPiinablau-def-1}
  \{\bTPiinablau\}_{ij}\,=\,
  \reallywidehat{\Pi_{{\rm i},ik}\partial_k u_{{\rm i},j}}
  -\widetilde{\Pi}_{{\rm i},ik}\partial_k \widehat{u}_{{\rm i},j}\,,
\end{equation}
%%%%%%%%%%%%%%%%%%%%%%%%%%%%%%%%%%%%%%%
%%%%%%%%%%%%%%%%%%%%%%%%%%%%%%%%%%%%%%%
\begin{equation}\label{eq:NHK_bTPiitimesB-def-1}
  \{\bTPiitimesB\}_{ij}\,=\,
  \reallywidehat{\Pi_{{\rm i},jl}\epsilon_{ilm}B_m}
  -\widetilde{\Pi}_{{\rm i},jl}\epsilon_{ilm}\widehat{B}_m\,.
\end{equation}
%%%%%%%%%%%%%%%%%%%%%%%%%%%%%%%%%%%%%%%
(In the above, sum over repeated indices is understood.)
Then, multiplying (\ref{eq:NHK_PressTens-filtered}) by $1/2$ and taking its trace (i.e., projecting onto $\frac{1}{2}\bb{I}$), one obtains~\citep{YangPOP2017}
%%%%%%%%%%%%%%%%%%%%%%%%%%%%%%%%%%%%%%%
\begin{equation}\label{eq:NHK_IonThermEn-filtered}
  \frac{\partial\,\whEPii}{\partial t}\, +\, 
  \nabla\cdot\Big(\widehat{{\cal E}_{\Pi_{\rm i}}\uiv}+\whqiv\Big)\, =\, 
  -\wtbPii:\nabla\whuiv\, -\,
  \TPiinablau\,,
\end{equation}
%%%%%%%%%%%%%%%%%%%%%%%%%%%%%%%%%%%%%%%
where $\TPiinablau$ is defined as the trace of $\bTPiinablau$,
%%%%%%%%%%%%%%%%%%%%%%%%%%%%%%%%%%%%%%%
\begin{equation}\label{eq:NHK_TPiinablau-def-1}
  \TPiinablau\,=\,
  \reallywidehat{\Pi_{{\rm i},jk}\partial_k u_{{\rm i},j}}
  -\widetilde{\Pi}_{{\rm i},jk}\partial_k \widehat{u}_{{\rm i},j}\,,
\end{equation}
%%%%%%%%%%%%%%%%%%%%%%%%%%%%%%%%%%%%%%%
and we have used the fact that the right-hand side of (\ref{eq:NHK_PressTens-filtered}) vanishes when taking its trace.

As we can see, the interaction between the pressure tensor and the strain tensor associated to scales $\geq\ell$, $-\widetilde{\bb{\Pi}}_{\mathrm{i},\ell}:\widehat{\bb{\Sigma}}_{\mathrm{i},\ell}$, is providing the (only) connection of the ion-thermal energy to another energy channel, that is with the ion-kinetic energy. 
At the same time, the (only) transfer mechanism of ion-thermal energy density through scale $\ell$ (i.e., the cascade rate of ion-internal energy) is provided by the ``turbulent'' pressure-strain non-linearity at scales $<\ell$, $\TPiinablau$.

\subsubsection{The filtered electric field and the role of sub-grid nonlinearities}\label{subsec:NHK_E-field-filtered}

By applying the Favre filter to the generalized Ohm's law (\ref{eq:NHK_Ohm}), we obtain%\footnote{Note that using electron continuity equation and the properties (\ref{eq:filter-def-2})--(\ref{eq:filter-def-3}) of the filters, one can show that the relation $\widehat{(\uev\cdot\grad)\uev}=\frac{1}{\wtvarrho}[\grad\cdot(\widetilde{\varrho\uev\uev})+\partial_t(\widetilde{\varrho\uev})]-\partial_t\whuev$ holds. Therefore, applying the Favre filter to equation (\ref{eq:NHK_Ohm}) is entirely equivalent to rewrite the electron momentum equation in its conservative form (see (\ref{app:eq:TF_Momentum})) and apply the regular filter (\ref{eq:filter-def-1}) on it.}
%%%%%%%%%%%%%%%%%%%%%%%%%%%%%%%%%%%%%%%
\begin{align}\label{eq:NHK_Ohm-filtered}
  \whEv\,= &
  \,-\frac{\whuev}{c}\times\whBv\,
  -\,\TuexB\,
  -\,\frac{\mi}{e}\frac{\grad\cdot\wtbPie}{\wtvarrho}\nonumber\\
  &\,-\,\frac{\me}{e}\frac{1}{\wtvarrho}\left[\grad\cdot\big(\wtvarrho\,\whuev\whuev+\Tueue\big)+\frac{\partial\,(\wtvarrho\,\whuev)}{\partial t}\right]\,,
\end{align}
%%%%%%%%%%%%%%%%%%%%%%%%%%%%%%%%%%%%%%%
where the sub-grid terms are now related to the electron-flow and $\uev\times\Bv$ non-linearities:
%%%%%%%%%%%%%%%%%%%%%%%%%%%%%%%%%%%%%%%
\begin{equation}\label{eq:NHK_Tueue-def-1}
  \Tueue\,=\,
  \widetilde{\varrho}\big(\widehat{\uev\uev}-\whuev\whuev\big)\,,
\end{equation}
%%%%%%%%%%%%%%%%%%%%%%%%%%%%%%%%%%%%%%%
%%%%%%%%%%%%%%%%%%%%%%%%%%%%%%%%%%%%%%%
\begin{equation}\label{eq:NHK_TuexB-def-1}
  \TuexB\,=\,\frac{1}{c}\Big(
  \reallywidehat{\uev\times\Bv}-\whuev\times\whBv\Big)\,.
\end{equation}
%%%%%%%%%%%%%%%%%%%%%%%%%%%%%%%%%%%%%%%
(Note that using electron continuity equation and the properties (\ref{eq:filter-def-2})--(\ref{eq:filter-def-3}) of the filters, one can show that the relation $\widehat{(\uev\cdot\grad)\uev}=\frac{1}{\wtvarrho}[\grad\cdot(\widetilde{\varrho\uev\uev})+\partial_t(\widetilde{\varrho\uev})]-\partial_t\whuev$ holds. Therefore, applying the Favre filter to equation (\ref{eq:NHK_Ohm}) is entirely equivalent to rewrite the electron momentum equation in its conservative form (see (\ref{app:eq:TF_Momentum})) and apply the regular filter (\ref{eq:filter-def-1}) on it.)
It is interesting to note that the filtered electric field at scales $\geq\ell$ in (\ref{eq:NHK_Ohm-filtered}) has an explicit contribution from the filtered fields at the same scales, plus a contribution $\bb{\epsilon}^*$ from scales $<\ell$,
\begin{widetext}
%%%%%%%%%%%%%%%%%%%%%%%%%%%%%%%%%%%%%%%
\begin{equation}\label{eq:NHK_E-filtered}
  \whEv\,=\,
  -\frac{\whuev}{c}\times\whBv\,
  -\,\frac{\mi}{e}\frac{\grad\cdot\wtbPie}{\wtvarrho}
  -\,\varepsilon_m\frac{\mi}{e}\frac{1}{\wtvarrho}\left[
  \grad\cdot\big(\wtvarrho\,\whuev\whuev\big)+\frac{\partial\,(\wtvarrho\,\whuev)}{\partial t}\right]\, 
  +\, \bb{\epsilon}^*\,,
\end{equation}
%%%%%%%%%%%%%%%%%%%%%%%%%%%%%%%%%%%%%%%
\end{widetext}
where we have defined $\bb{\epsilon}^*$ as the sub-grid electric field,
%%%%%%%%%%%%%%%%%%%%%%%%%%%%%%%%%%%%%%%
\begin{equation}\label{eq:NHK_E-subgrid}
  \bb{\epsilon}^*\,\equiv\,
  -\TuexB\,-\,
  \varepsilon_m\frac{\mi}{e}\frac{1}{\wtvarrho}\grad\cdot\Tueue\,,
\end{equation}
%%%%%%%%%%%%%%%%%%%%%%%%%%%%%%%%%%%%%%%
sometimes referred to as ``turbulent'' electric field, which is arising from unresolved scales $<\ell$ due to the nonlinear contributions.
It is worth to further decompose this sub-grid electric field into its ``MHD contribution'' (i.e., due to the $\uiv\times\Bv$ nonlinearities), ``Hall contribution'' (i.e., related to the $\Jv\times\Bv$ term), and ``electron-inertia contribution'' (i.e., associated with the $\uev\uev$ nonlinearity that is retained when $\epsilon_m\neq0$),
\[
\bb{\epsilon}^*\,=\,
\bb{\epsilon}_{\mathrm{MHD}}^*\,
+\,\bb{\epsilon}_{\mathrm{Hall}}^*\, 
+\,\bb{\epsilon}_{d_\mathrm{e}}^*\,,
\]
with
%%%%%%%%%%%%%%%%%%%%%%%%%%%%%%%%%%%%%%%
\begin{align}
  \bb{\epsilon}_{\mathrm{MHD}}^*\,=\, & -\TuxB\label{eq:NHK_Emhd-subgrid}\\
  \bb{\epsilon}_{\mathrm{Hall}}^*\,=\, & -\TJxB\label{eq:NHK_Ehall-subgrid}\\
  \bb{\epsilon}_{d_\mathrm{e}}^*\,=\, & -\varepsilon_m\frac{\mi}{e}\frac{1}{\wtvarrho}\grad\cdot\Tueue\label{eq:NHK_Ede-subgrid}\,,
\end{align}
%%%%%%%%%%%%%%%%%%%%%%%%%%%%%%%%%%%%%%%
where we have used the identity $\TuexB=\TuxB+\TJxB$ in which the $\Jv\times\Bv$ sub-grid term is defined as
%%%%%%%%%%%%%%%%%%%%%%%%%%%%%%%%%%%%%%%
\begin{equation}\label{eq:NHK_TJxB-def-1}
  \TJxB\,\equiv\,\frac{\mi}{ec}\frac{1}{\wtvarrho}\left(
  \reallywidetildeB{\Jv\times\Bv}-\widetilde{\Jv}\times\widetilde{\Bv}\right)\,.
\end{equation}
%%%%%%%%%%%%%%%%%%%%%%%%%%%%%%%%%%%%%%%
At this point we can clearly see the difference between an MHD treatment~\citep{AluieNJP2017} and the hybrid-kinetic case. In fact, in the coarse-grained MHD equations one would only obtain the sub-grid electromotive force associated to the $\uiv\btimes\Bv$ fluctuations, equation (\ref{eq:NHK_Emhd-subgrid}). As soon as we allow for a two-fluid-like description, the additional terms in equations (\ref{eq:NHK_Ehall-subgrid}) and (\ref{eq:NHK_Ede-subgrid}) arise. In particular, the Hall sub-grid electric field, $\bb{\epsilon}_{\mathrm{Hall}}^*$, accounts for the separation between the ion dynamics and the magnetic-field dynamics (which is now frozen into the electron-fluid dynamics) that occurs as the ion-inertial scales are approached. This term should actually be captured also in the Hall-MHD limit. The last term, $\bb{\epsilon}_{d_\mathrm{e}}^*$, arises from finite-inertia effects of the electron fluid, $\varepsilon_m\neq0$, which now contributes to the sub-grid electric field through (the divergence of) its sub-grid Reynolds stress, $\Tueue$. Therefore, this term could not be captured by a MHD or Hall-MHD treatment.
Note that in the massless-electron limit, the only contributions to the ``turbulent'' electric field comes from the MHD and Hall terms, $\bb{\epsilon}_{\mathrm{MHD}}^*$ and $\bb{\epsilon}_{\mathrm{Hall}}^*$, respectively. As we will see later, in Section~\ref{subsec:NHKtotalenergy_filtered}, this has a direct implication on the turbulent energy transfer across scales. In particular, this formalism seems to provide a proof of the assumption made in \citet{HellingerAPJL2018} that in a hybrid model with isothermal, massless electrons this transfer can be approximated by its additive incompressive contributions due to the MHD and Hall terms.
If one is interested to investigate in more details the mechanisms underlying the electron-inertia term, $\bb{\epsilon}_{d_\mathrm{e}}^*$, it could be further decomposed by using the relation $\Tueue=\Tuu+\TJJ-\TJuuJ$, i.e.
%%%%%%%%%%%%%%%%%%%%%%%%%%%%%%%%%%%%%%%
\begin{equation}\label{eq:NHK_Ede-subgrid-2}
  \bb{\epsilon}_{d_\mathrm{e}}^*\,=\,  -\varepsilon_m\frac{\mi}{e}\frac{1}{\wtvarrho}\grad\cdot\big(\Tuu+\TJJ-\TJuuJ\big)\,,
\end{equation}
%%%%%%%%%%%%%%%%%%%%%%%%%%%%%%%%%%%%%%%
where the ``current-current'' and ``current-flow'' sub-grid terms are given by
%%%%%%%%%%%%%%%%%%%%%%%%%%%%%%%%%%%%%%%
\begin{equation}\label{eq:NHK_TJJ-def-1}
  \TJJ\,\equiv\,
  \frac{\mi^2}{e^2}\left(
  \widetilde{\frac{\Jv\Jv}{\varrho}}-\frac{\widetilde{\Jv}\widetilde{\Jv}}{\wtvarrho}\right)\,,
\end{equation}
%%%%%%%%%%%%%%%%%%%%%%%%%%%%%%%%%%%%%%%
%%%%%%%%%%%%%%%%%%%%%%%%%%%%%%%%%%%%%%%
\begin{equation}\label{eq:NHK_TJu-def-1}
  \TJuuJ\,\equiv\,
  \frac{\mi}{e}\left[
  (\widetilde{\Jv\uiv}+\widetilde{\uiv\Jv})-(\widetilde{\Jv}\widetilde{\uv}_\mathrm{i}+\widetilde{\uv}_\mathrm{i}\widetilde{\Jv})\right]\,.
\end{equation}
%%%%%%%%%%%%%%%%%%%%%%%%%%%%%%%%%%%%%%%
and they can be seen as turbulent Reynolds stresses associated to the ``current-current'' and ``current-flow'' non-linearities that add to the turbulent Reynolds stress of the ion flow.

\subsubsection{Filtered electron bulk (kinetic) energy}

By taking the scalar product of (\ref{eq:NHK_Ohm-filtered}) with $\frac{e}{\mi}\wtvarrho\whuev$, one obtains the filtered equation for the equivalent electron bulk energy density:
%%%%%%%%%%%%%%%%%%%%%%%%%%%%%%%%%%%%%%%
\begin{align}\label{eq:NHK_ElectronKinEn-filtered}
  \frac{\partial\,\whEue}{\partial t}\, +\, 
  \grad\cdot\big(\whEue\whuev\big)\, =\, &
  -\,\whuev\cdot\Big[\grad\cdot\big(\wtbPie+\varepsilon_m\Tueue\big)\Big]\nonumber\\
 & \, -\frac{e}{\mi}\wtvarrho\,\whuev\cdot\big(\whEv+\TuexB\big)\,,
\end{align}
%%%%%%%%%%%%%%%%%%%%%%%%%%%%%%%%%%%%%%%
or, equivalently, as
%%%%%%%%%%%%%%%%%%%%%%%%%%%%%%%%%%%%%%%
\begin{equation}\label{eq:NHK_ElectronKinEn-filtered-2}
  \frac{\partial\,\whEue}{\partial t}\, +\, 
  \grad\cdot\big(\whEue\whuev\,+\,\wtbPie\cdot\whuev\big)\,=\, \,\wtbPie:\grad\whuev\,
  +\,\widehat{\bb{j}}_\mathrm{e}\cdot(\whEv\,
  -\,\bb{\epsilon}^*)\,,
\end{equation}
%%%%%%%%%%%%%%%%%%%%%%%%%%%%%%%%%%%%%%%
where the filtered electron bulk energy density is defined as $\whEue=\varepsilon_m\frac{1}{2}\wtvarrho|\whuev|^2$
and we have introduced the Favre-filtered electron current density, $\widehat{\bb{j}}_\mathrm{e}\equiv-\frac{e}{\mi}\wtvarrho\,\whuev=-e\widetilde{n}\whuev$.
Therefore, when considering a characteristic scale $\ell$ for the filters, the energy transfer of electrons' kinetic energy through that scale is mediated by the interaction between the electron current density at scales $\geq\ell$, $\widehat{\bb{j}}_{\mathrm{e},\ell}$, and the ``turbulent'' electric-field fluctuations at scales $<\ell$, $\bb{\epsilon}_\ell^*$, that is $\mathcal{T}_{\mathrm{kin},\ell}^{(\mathrm{e})}=-\widehat{\bb{j}}_{\mathrm{e},\ell}\cdot\bb{\epsilon}_\ell^*$.
Also note that it would be possible to rewrite (\ref{eq:NHK_ElectronKinEn-filtered-2}) in an analogous way as done for (\ref{eq:NHK_IonKinEn-filtered-1bis}), so that the energy flux through scale $\ell$ is given by the combination of the electrons' current density and the electrons' strain tensor at scales $\geq\ell$ interacting with the turbulent ``MHD+Hall'' electric field and with the turbulent electron-flow Reynolds stress, respectively, i.e. $\mathcal{T}_{\mathrm{kin},\ell}^{(\mathrm{e})}=-\widehat{\bb{j}}_{\mathrm{e},\ell}\cdot(\bb{\epsilon}_{\mathrm{MHD},\ell}^*+\bb{\epsilon}_{\mathrm{Hall},\ell}^*)+\varepsilon_m\bb{\mathcal{T}}_{uu,\ell}^{(\mathrm{e})}:\widehat{\bb{\Sigma}}_{\mathrm{e},\ell}$.
The other two terms instead represent the connection between the electrons' kinetic energy channel and the other energy channels at scales $\geq\ell$, i.e., to the magnetic-field energy density (through the current-field interaction, $\widehat{\bb{j}}_{\mathrm{e},\ell}\cdot\whEv_\ell$) and to the electrons' thermal energy density (through the pressure-strain interaction, $\widetilde{\bb{\Pi}}_{\mathrm{e},\ell}:\widehat{\bb{\Sigma}}_{\mathrm{e},\ell}$).

In the limit of massless electrons, $\varepsilon_m\to0$, equation (\ref{eq:NHK_ElectronKinEn-filtered}) reduces to the filtered version of the balance equation in (\ref{eq:NHK_ElectronKinEn-massless}),
%%%%%%%%%%%%%%%%%%%%%%%%%%%%%%%%%%%%%%%
\begin{equation}\label{eq:NHK_ElectronKinEn-filtered-massless}
  \frac{e}{\mi}\wtvarrho\,\whuev\cdot\left(\whEv\,
  +\,\frac{\mi}{e}\frac{\grad\cdot\wtbPie}{\wtvarrho}\right)\,= \frac{e}{\mi}\wtvarrho\,\whuev\cdot\bb{\epsilon}^*\,,
\end{equation}
%%%%%%%%%%%%%%%%%%%%%%%%%%%%%%%%%%%%%%%
where now $\bb{\epsilon}^*=-\TuexB=\bb{\epsilon}_\mathrm{MHD}^*+\bb{\epsilon}_\mathrm{Hall}^*$. Note that this equation, when space averaged, provides a sort of ``vertex-conservation law'' through the ``disappearing'' electrons' kinetic energy channel in the limit $\varepsilon_m\to0$.
Note that, as discussed in the previous Section, even in this limit there is a difference between the scale-to-scale transfer that is found in coarse-grained MHD equations~\citep{AluieNJP2017,BianAluiePRL2019} and the one operating in hybrid models, which arises from the Hall term. Interestingly, a similar difference in the sub-grid turbulent electric field, and thus in the associated scale-to-scale magnetic-energy transfer, can be found from Hall-MHD equations. This may indeed explain why Hall-MHD and hybrid-kinetics with isothermal, massless electrons seem to provide similar results in terms of the turbulent cascade of magnetic-field fluctuations (see, e.g., \citet{PapiniAPJ2019}) and why this transfer may be approximated by the additive contributions due to the MHD and Hall terms in the sub-grid electric field (see, e.g., \citet{HellingerAPJL2018}; see also the discussion associated to Figure~\ref{fig:NHKenergychannels_masslessisothermal}). Indeed, the relevance and nature of the Hall electric-field fluctuations in possibly mediating the transition to a so-called reconnection-mediated regime~\citep{CerriCalifanoNJP2017,FranciAPJL2017,LoureiroBoldyrevAPJ2017,MalletJPP2017} of sub-ion-scale turbulent energy transfer was already pointed out by means of hybrid-kinetic simulations in \citet{CerriCalifanoNJP2017}.

\subsubsection{Filtered electron inetrnal (thermal) energy}

We now apply the Favre filter also to the electron pressure equations, (\ref{eq:NHK_Pe-polytropic-2}) or (\ref{eq:NHK_PressTens-electrons}) depending on the closure adopted:\\

\begin{itemize}
    \item[(a)] {\em isotropic, polytropic fluid}.
    %%%%%%%%%%%%%%%%%%%%%%%%%%%%%%%%%%%%%%%
    \begin{equation}\label{eq:NHK_Pe-polytropic-2-filtered}
     \frac{\partial\,\whPe}{\partial t}\,
     +\,\grad\cdot\big(\widehat{P_{\rm e}\uev}\big)\, =\, (1-\gamma)\Big[\wtPe\big(\grad\cdot\whuev\big)
     +\TPenablaue\Big]\,,
    \end{equation}
    %%%%%%%%%%%%%%%%%%%%%%%%%%%%%%%%%%%%%%%
    where $\TPenablaue$ is the sub-grid term associated to the isotropic-compression contribution to the pressure-strain electrons' non-linearity, defined as
    %%%%%%%%%%%%%%%%%%%%%%%%%%%%%%%%%%%%%%%
    \begin{equation}\label{eq:NHK_TPenablaue-def-1}
    \TPenablaue\,=\,
    \reallywidehat{P_{\rm e}(\grad\cdot\uev)}
    -\wtPe(\grad\cdot\whuev)\,.
    \end{equation}
    %%%%%%%%%%%%%%%%%%%%%%%%%%%%%%%%%%%%%%%
    
    \item[(b)] {\em fully anisotropic, adiabatic fluid}. 
    %%%%%%%%%%%%%%%%%%%%%%%%%%%%%%%%%%%%%%%
    \begin{align}
     \frac{\partial\,\whbPie}{\partial t} 
     + \grad\cdot\big(\widehat{\bPie\uev}\big) = 
     & \,-\Big\{\big(\wtbPie\cdot\grad\big)\whuev+\bTPienablaue\Big\}^{\rm sym}\nonumber\\
     & \,+\frac{e}{\me c}\Big\{\wtbPie\times\whBv+\bTPietimesB\Big\}^{\rm sym},\label{eq:NHK_ElectronPressTens-filtered}
    \end{align}
    %%%%%%%%%%%%%%%%%%%%%%%%%%%%%%%%%%%%%%%
    where we have introduced the electrons counterpart of the sub-grid terms related to the pressure-strain and pressure-magnetic tensor non-linearity:
    %%%%%%%%%%%%%%%%%%%%%%%%%%%%%%%%%%%%%%% 
    \begin{equation}\label{eq:NHK_bTPienablaue-def-1}
     \{\bTPienablaue\}_{ij}\,=\,
     \reallywidehat{\Pi_{{\rm e},ik}\partial_k u_{{\rm e},j}}
     -\widetilde{\Pi}_{{\rm e},ik}\partial_k \widehat{u}_{{\rm e},j}\,,
    \end{equation}
    %%%%%%%%%%%%%%%%%%%%%%%%%%%%%%%%%%%%%%%
    %%%%%%%%%%%%%%%%%%%%%%%%%%%%%%%%%%%%%%%
    \begin{equation}\label{eq:NHK_bTPietimesB-def-1}
     \{\bTPietimesB\}_{ij}\,=\,
     \epsilon_{ilm}\big(\reallywidehat{\Pi_{{\rm e},jl}B_m}
     -\widetilde{\Pi}_{{\rm e},jl}\widehat{B}_m\big)\,.
    \end{equation}
    %%%%%%%%%%%%%%%%%%%%%%%%%%%%%%%%%%%%%%%
    
\end{itemize}
From the above equations, the correspondent equations for the filtered electron internal energy density follow:
%%%%%%%%%%%%%%%%%%%%%%%%%%%%%%%%%%%%%%%
\begin{equation}\label{eq:NHK_ElectronThermEn-filtered}
  \frac{\partial\,\whEPie}{\partial t}\, 
  +\, \grad\cdot\big(\widehat{{\cal E}_{\Pi_{\rm e}}\uev}\big)\, 
  =\, \left\{\begin{array}{l}
       (1-\gamma)\frac{3}{2}\Big[\wtPe\big(\grad\cdot\whuev\big)
       \,+\TPenablaue\Big]\\
       \,\\
      -\wtbPie:\grad\whuev\,
      -\,\TPienablaue
  \end{array}\right.\begin{array}{c}
        \\%\qquad\textrm{(a)}\\
        \, \\
        %\qquad\textrm{(b)}
  \end{array}\,,
\end{equation}
%%%%%%%%%%%%%%%%%%%%%%%%%%%%%%%%%%%%%%%
where $\whEPie=\frac{1}{2}{\rm tr}[\whbPie]$ and the sub-grid term on the right-hand side of (\ref{eq:NHK_ElectronThermEn-filtered}--b) is defined as the trace of $\bTPienablaue$ given in (\ref{eq:NHK_bTPienablaue-def-1}), i.e., $\TPienablaue={\rm tr}[\bTPienablaue]$.

\subsubsection{Filtered magnetic energy}

Noticing that the Favre-filtered Maxwell equations of the NHK model are
%%%%%%%%%%%%%%%%%%%%%%%%%%%%%%%%%%%%%%%
\begin{equation}\label{eq:NHK_Maxwell-filtered}
\frac{\partial\,\whBv}{\partial t}\, =\, -\,c\,\grad\times\whEv\,,\quad
\whJv\,=\,\frac{c}{4\pi}\,\grad\times\whBv
\end{equation}
%%%%%%%%%%%%%%%%%%%%%%%%%%%%%%%%%%%%%%%
the equation for the filtered magnetic energy, $\whEB=|\whBv|^2/8\pi$, is readily derived in the usual way:
%%%%%%%%%%%%%%%%%%%%%%%%%%%%%%%%%%%%%%%
\begin{align}
  \frac{\partial\,\whEB}{\partial t}\, 
  +\, \grad\cdot\left(\frac{\whEv\times\whBv}{4\pi}c\right)\, 
  = & \, -\whJv\cdot\whEv\nonumber\\
  =\, &
  \,-\,\frac{e}{\mi}\wtvarrho(\whuiv-\whuev)\cdot\whEv\,
  -\,\TJE\,,\label{eq:NHK_MagneticEn-filtered}
\end{align}
%%%%%%%%%%%%%%%%%%%%%%%%%%%%%%%%%%%%%%%
where the sub-grid term associated with wave-particle interaction is defined as%\footnote{The actual sub-grid non-linearity here is given by $\Tnua=\widehat{n\uav}-\widehat{n}\whuav$, so that the decomposition $\whJv=e\widehat{n}(\whuiv-\whuev)+\Tnu-\Tnue$ holds. The resulting sub-grid contribution to the r.h.s. of equation (\ref{eq:NHK_MagneticEn-filtered}) would then be $-(\Tnu-\Tnue)\cdot\whEv$, that we have called $-\TJE$ for shortness.}
%%%%%%%%%%%%%%%%%%%%%%%%%%%%%%%%%%%%%%%
\begin{align}\label{eq:NHK_TJE-def-1}
  \TJE\,= 
  & \,\left[\sum_\alpha e_\alpha(\widehat{n\uav}-\widetilde{n}\widehat{\bb{u}}_\alpha) \right]\cdot\whEv\nonumber\\
  = 
  &\,\frac{e}{\mi}\Big[(\widehat{\varrho\uiv}-\widehat{\varrho\uev})
  -\wtvarrho\,(\whuiv-\whuev)\Big]\cdot\whEv\,.
\end{align}
%%%%%%%%%%%%%%%%%%%%%%%%%%%%%%%%%%%%%%%
(The actual sub-grid non-linearity here is given by $\Tnua=\widehat{n\uav}-\widehat{n}\whuav$, so that the decomposition $\whJv=e\widehat{n}(\whuiv-\whuev)+\Tnu-\Tnue$ holds. The resulting sub-grid contribution to the r.h.s. of equation (\ref{eq:NHK_MagneticEn-filtered}) would then be $-(\Tnu-\Tnue)\cdot\whEv$, that we have called $-\TJE$ for shortness.)

Analogously to what has been done for the electric field in (\ref{eq:NHK_E-filtered}), we can define a sub-grid or ``turbulent'' current density,
%%%%%%%%%%%%%%%%%%%%%%%%%%%%%%%%%%%%%%%
\begin{equation}\label{eq:NHK_J-subgrid}
  \jv^*\,\equiv\,
  \frac{1}{\wtvarrho}\left(\widetilde{\varrho\Jv}-\wtvarrho\,\widetilde{\Jv}\right)=\Tnu-\Tnue\,,
\end{equation}
%%%%%%%%%%%%%%%%%%%%%%%%%%%%%%%%%%%%%%%
so that $\whJv=\widetilde{\Jv}+\jv^*=\widehat{\jv}_\mathrm{i}+\widehat{\jv}_\mathrm{e}+\jv^*$ (note the different filters on $\Jv$),
%%%%%%%%%%%%%%%%%%%%%%%%%%%%%%%%%%%%%%%
\begin{equation}\label{eq:NHK_MagneticEn-filtered-2}
  \frac{\partial\,\whEB}{\partial t}\, 
  +\, \grad\cdot\left(\frac{\whEv\times\whBv}{4\pi}c\right)\, 
  =\,
  -\,(\widetilde{\Jv}\,
  +\jv^*)\cdot\whEv\,,
\end{equation}
%%%%%%%%%%%%%%%%%%%%%%%%%%%%%%%%%%%%%%%
and, if $\ell$ is the characteristic filtering scale, we can interpret the energy transfer of magnetic energy through that scale as the result of the interaction between the ``turbulent'' currents at scales $<\ell$, $\bb{j}_\ell^*$, and the electric field at scales $\geq\ell$, $\whEv_\ell$, i.e., $\mathcal{T}_\ell^{(\mathrm{mag})}=-\bb{j}_\ell^*\cdot\whEv_\ell=-\TJE$.
On the other hand, the term $-\widetilde{\Jv}_\ell\cdot\whEv_\ell=-(\sum_\alpha\widehat{\bb{j}}_{\alpha,\ell})\cdot\whEv_\ell$ represent the energy density flux within scales $\geq\ell$ that couples the magnetic energy density channel to the species' kinetic energy densities.

\subsection{The filtered equation for the total energy density in NHK}\label{subsec:NHKtotalenergy_filtered}

Gathering all the previous equations for the filtering energy densities (and just rewriting few terms in a convenient way), they read:
\begin{widetext}
%%%%%%%%%%%%%%%%%%%%%%%%%%%%%%%%%%%%%%%
\begin{align}
    \frac{\partial\,\whEu}{\partial t} + 
  \grad\cdot\Big[\whEu\whuiv+\big(\wtbPii+\Tuu\big)\cdot\whuiv\Big] =\, &
  \,\wtvarrho\,\whuiv\cdot\whFextv\,
  +\,\widehat{\jv}_\mathrm{i}\cdot\big(\whEv\,-\,\bb{\epsilon}_\mathrm{MHD}^*\big)\, +\big(\wtbPii\,+\,\Tuu\big):\grad\whuiv\,,\label{eq:NHK_energy-summary-ionkin-filtered}\\
  %%%%
    \frac{\partial\,\whEPii}{\partial t} + 
  \grad\cdot\Big(\widehat{{\cal E}_{\Pi_{\rm i}}\uiv}+\whqiv\Big)=\, & 
  \,-\wtbPii:\grad\whuiv\,
  -\,\TPiinablau\,,\label{eq:NHK_energy-summary-iontherm-filtered}\\
  %%%%
  \frac{\partial\,\whEue}{\partial t} + 
  \grad\cdot\Big[\whEue\whuev+
  \big(\wtbPie+\varepsilon_m\Tueue\big)\cdot\whuev\Big] =\, &
  \,\widehat{\jv}_\mathrm{e}\cdot\big(\whEv\,\,-\,\bb{\epsilon}_\mathrm{MHD}^*\,-\,\bb{\epsilon}_\mathrm{Hall}^*\big)\, +\,\big(\wtbPie+\varepsilon_m\Tueue\big):\grad\whuev\,,
\label{eq:NHK_energy-summary-eleckin-filtered}\\
  %%%%
    \hspace{-5cm}\frac{\partial\,\whEPie}{\partial t} 
  + \grad\cdot\big(\widehat{{\cal E}_{\Pi_{\rm e}}\uev}\big) 
  =\,& \left\{\begin{array}{l}
       (1-\gamma)\frac{3}{2}\Big[\wtPe\big(\grad\cdot\whuev\big)
       \,+\TPenablaue\Big]\\
       \,\\
      -\wtbPie:\grad\whuev\,
      -\,\TPienablaue
  \end{array}\right.\begin{array}{c}
        \textrm{(a)}\\
        \, \\
        \textrm{(b)}
  \end{array}\,,\label{eq:NHK_energy-summary-electherm-filtered}\\
  %%%%
  \frac{\partial\,\whEB}{\partial t} 
  + \grad\cdot\left(\frac{\whEv\times\whBv}{4\pi}c\right) 
  =\, & 
  \,-\,\big(\widehat{\jv}_\mathrm{i}\,
  +\,\widehat{\jv}_\mathrm{e}\,+\,\jv^*\big)\cdot\whEv\label{eq:NHK_energy-summary-magn-filtered}\,,
\end{align}
%%%%%%%%%%%%%%%%%%%%%%%%%%%%%%%%%%%%%%%
\end{widetext}
where we remind that $\widehat{\jv}_\alpha\equiv\frac{e_\alpha}{\mi}\wtvarrho\,\whuav=e_\alpha\widetilde{n}\whuav$ is the species' current density (i.e., carrying the appropriate sign due to the species' charge within its definition).
Then, summing up the above equations one obtains the equation for the total energy density:
%%%%%%%%%%%%%%%%%%%%%%%%%%%%%%%%%%%%%%%
\begin{equation}\label{eq:NHK_TotalEn-filtered}
  \frac{\partial\,\widehat{\cal E}}{\partial t}\, 
  +\,\grad\cdot\widehat{\bb{\Phi}}_{\widehat{\cal E}}\, 
  =\,\wtvarrho\,\whuiv\cdot\whFextv\,
  +\,\widehat{\cal I}_{\rm e}\,
  +\,\whcalSsg[0]\,
  +\,\whcalSsg[m_{\rm e}]\,,
\end{equation}
%%%%%%%%%%%%%%%%%%%%%%%%%%%%%%%%%%%%%%%
with the filtered total energy density and energy density flux defined as 
%%%%%%%%%%%%%%%%%%%%%%%%%%%%%%%%%%%%%%%
\begin{equation}\label{eq:NHK_TotalEnDens-filtered}
  \widehat{\cal E}\,=\,
  \whEu\,+\,\whEPii\,
  +\,\whEue\,+\,\whEPie\,
  +\,\whEB\,,
\end{equation}
%%%%%%%%%%%%%%%%%%%%%%%%%%%%%%%%%%%%%%%
%%%%%%%%%%%%%%%%%%%%%%%%%%%%%%%%%%%%%%%
\begin{align}
  \widehat{\bb{\Phi}}_{\widehat{\cal E}}\,=\,
  & \,\,\whEu\whuiv\,
  +\,\widehat{{\cal E}_{\Pi_{\rm i}}\uiv}\,
  +\,(\wtbPii+\Tuu)\cdot\whuiv\,
  +\,\whqiv\nonumber\\
  &\,+\,\whEue\whuev\,
  +\,\widehat{{\cal E}_{\Pi_{\rm e}}\uev}\,
  +\,(\wtbPie+\varepsilon_m\Tueue)\cdot\whuev\label{eq:NHK_TotalEnFlux-filtered}\\
  &\,+\,\frac{\whEv\times\whBv}{4\pi}c\,,\nonumber
\end{align}
%%%%%%%%%%%%%%%%%%%%%%%%%%%%%%%%%%%%%%%
respectively.
With respect to its non-filtered version in (\ref{eq:NHK_TotalEn-1}), the filtered equation for the total energy density shows additional terms.
In fact, on the right-hand side of (\ref{eq:NHK_TotalEn-filtered}), in addition to the filtered version of the forcing injection, $\wtvarrho\,\whuiv\cdot\whFextv$, and of the term taking into account for the electron closure,
%%%%%%%%%%%%%%%%%%%%%%%%%%%%%%%%%%%%%%%
\begin{equation}\label{eq:NHK_Ie-term-filtered}
  \widehat{\cal I}_{\rm e}\, 
  =\, 
  \left\{\begin{array}{l}
    \left(\frac{5}{3}-\gamma\right)\frac{3}{2}\widetilde{P}_{\rm e}(\grad\cdot\whuev)\\

      \\  

    0    
  \end{array}\right.\begin{array}{c}
        \qquad\textrm{(a)}\\
        \, \\
        \qquad\textrm{(b)}
  \end{array}\,,
\end{equation}
%%%%%%%%%%%%%%%%%%%%%%%%%%%%%%%%%%%%%%%
now there are ``source-like'' (or ``sink-like'') sub-grid term, $\whcalSsgtot$, defined as
%%%%%%%%%%%%%%%%%%%%%%%%%%%%%%%%%%%%%%%
\begin{equation}
  \whcalSsg[0]\,=\,
  \widehat{\jv}_\mathrm{i}\cdot\TuxB\,
  +\,\Tuu:\nabla\whuiv\,
  -\,\TPiinablau\,
  -\,\TJE\,
  +\,\widehat{\jv}_\mathrm{e}\cdot\TuexB
  \,-\,\TPienablaue\,,\label{eq:NHK_whcalSsg-0}
\end{equation}
%%%%%%%%%%%%%%%%%%%%%%%%%%%%%%%%%%%%%%%
%%%%%%%%%%%%%%%%%%%%%%%%%%%%%%%%%%%%%%%
\begin{equation}
  \whcalSsg[m_{\rm e}]\,=\,
  \varepsilon_m\Tueue:\nabla\whuev\,.\label{eq:NHK_whcalSsg-me}
\end{equation}
%%%%%%%%%%%%%%%%%%%%%%%%%%%%%%%%%%%%%%%
where we have separated the base NHK contribution, $\whcalSsg[0]$, from the one arising from finite-electron-inertia effects, $\whcalSsg[m_{\rm e}]$.
Notice that above we have used $\TPienablaue$ for shortness, but its definition actually depends on the closure adopted for the electron fluid (cf. equations (\ref{eq:NHK_TPenablaue-def-1}), (\ref{eq:NHK_bTPienablaue-def-1}) and (\ref{eq:NHK_ElectronThermEn-filtered})),
%%%%%%%%%%%%%%%%%%%%%%%%%%%%%%%%%%%%%%%
\begin{equation}\label{eq:NHK_TPienablaue-split}
  \TPienablaue\, 
  =\, 
  \left\{\begin{array}{l}
    \left(1-\gamma\right)\frac{3}{2}
    \left[\reallywidehat{P_{\rm e}(\grad\cdot\uev)}
    -\widetilde{P}_{\rm e}(\grad\cdot\whuev)\right]\\
    \,  \\  
    \reallywidehat{\bPie:\grad\uev}-\wtbPie:\grad\whuev    
  \end{array}\right.\begin{array}{c}
        \qquad\textrm{(a)}\\
        \, \\
        \qquad\textrm{(b)}
  \end{array}\,,
\end{equation}
%%%%%%%%%%%%%%%%%%%%%%%%%%%%%%%%%%%%%%%
whereas the definition of the other sub-grid terms ${\cal T}$, that we report here for completeness, is the same as given above:
%%%%%%%%%%%%%%%%%%%%%%%%%%%%%%%%%%%%%%%
\begin{equation}\label{eq:NHK_Tuaua-def-1}
  \Tuaua\,=\,
  \widetilde{\varrho}\big(\reallywidehat{\uav\uav}\,
  -\,\whuav\whuav\big)\,,
\end{equation}
%%%%%%%%%%%%%%%%%%%%%%%%%%%%%%%%%%%%%%%
%%%%%%%%%%%%%%%%%%%%%%%%%%%%%%%%%%%%%%%
\begin{equation}\label{eq:NHK_TuxB-def-1a}
  \TuxB\,=\,\frac{1}{c}\Big(
  \reallywidehat{\uiv\times\Bv}\,
  -\,\whuiv\times\whBv\Big)\,=\,-\bb{\epsilon}_\mathrm{MHD}^*\,.
\end{equation}
%%%%%%%%%%%%%%%%%%%%%%%%%%%%%%%%%%%%%%%
%%%%%%%%%%%%%%%%%%%%%%%%%%%%%%%%%%%%%%%
\begin{equation}\label{eq:NHK_TuxB-def-1b}
  \TuexB\,=\,\frac{1}{c}\Big(
  \reallywidehat{\uev\times\Bv}\,
  -\,\whuev\times\whBv\Big)\,=\,-(\bb{\epsilon}_\mathrm{MHD}^*+\bb{\epsilon}_\mathrm{Hall}^*)\,.
\end{equation}
%%%%%%%%%%%%%%%%%%%%%%%%%%%%%%%%%%%%%%%
%%%%%%%%%%%%%%%%%%%%%%%%%%%%%%%%%%%%%%%
\begin{equation}\label{eq:NHK_TPiinablau-def-2}
  \TPiinablau\,=\,
  \reallywidehat{\bPii:\grad\uiv}\,
  -\,\wtbPii:\grad\whuiv\,,
\end{equation}
%%%%%%%%%%%%%%%%%%%%%%%%%%%%%%%%%%%%%%%
%%%%%%%%%%%%%%%%%%%%%%%%%%%%%%%%%%%%%%%
\begin{equation}\label{eq:NHK_TJE-def-2}
  \TJE\,=\,
  \,\frac{e}{\mi}\Big[(\widehat{\varrho\uiv}-\wtvarrho\,\whuiv)
  -\,(\widehat{\varrho\uev}-\wtvarrho\,\whuev)\Big]\cdot\whEv\,=\,\jv^*\cdot\whEv\,.
\end{equation}
%%%%%%%%%%%%%%%%%%%%%%%%%%%%%%%%%%%%%%%

\begin{figure*}
    %\centering
    \includegraphics[width=\columnwidth]{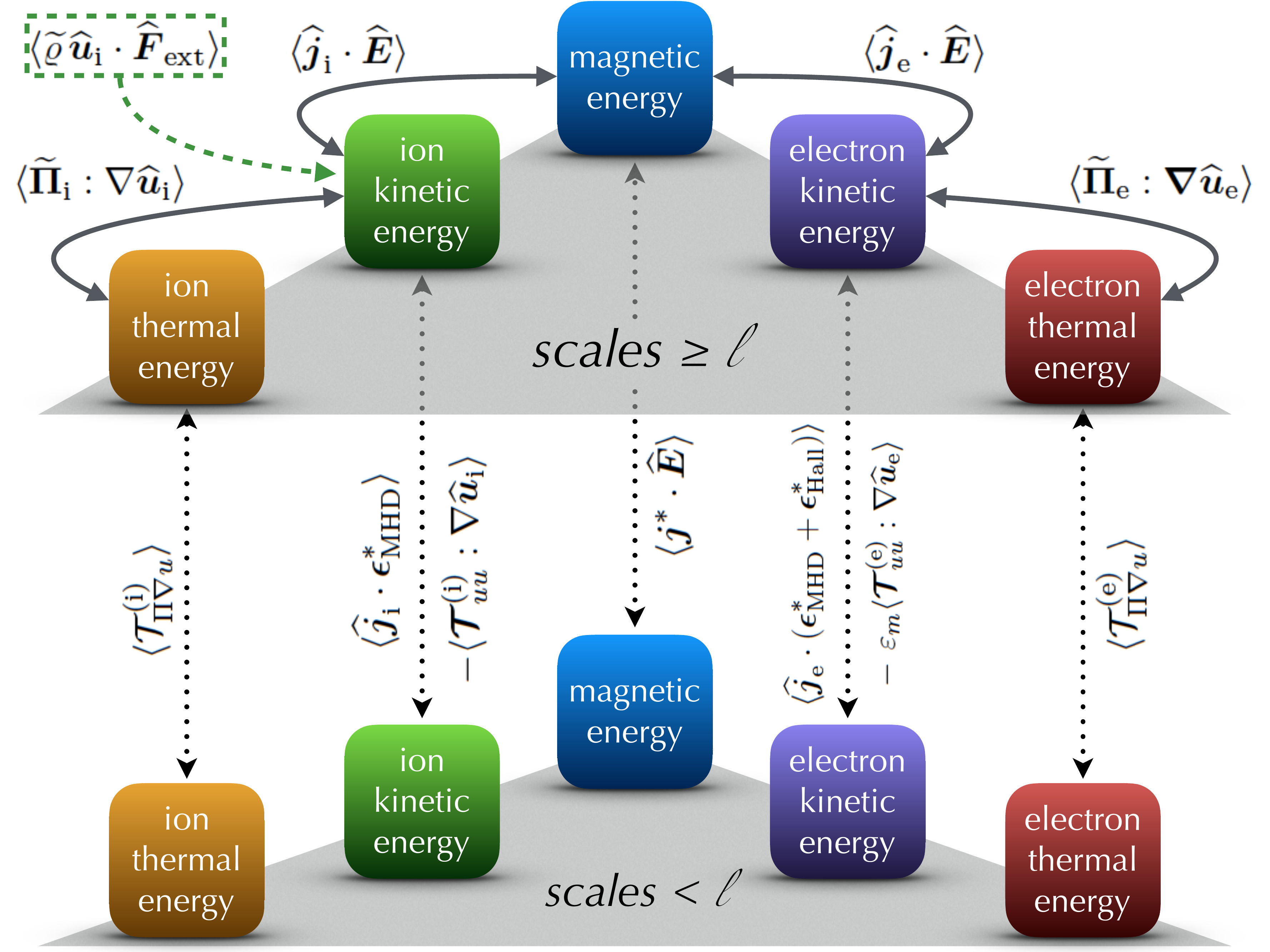}%
    \caption{Global energy channels of the NHK model: schematic cartoon of the space-averaged filtered energy equations in (\ref{eq:NHK_energy-summary-ionkin-filtered-avg})--(\ref{eq:NHK_energy-summary-magn-filtered-avg}). The scale $\ell$ denotes the filter's characteristic scale.}
    \label{fig:NHKenergychannels}
\end{figure*}

Analogously to what was done in equations (\ref{eq:NHK_energy-summary-ionkin})--(\ref{eq:NHK_energy-summary-magn}), it may be interesting to look at the space-averaged version of the filtered equations (\ref{eq:NHK_energy-summary-ionkin-filtered})--(\ref{eq:NHK_energy-summary-magn-filtered}).
By taking such space average and assuming again vanishing fluxes at the boundaries, one obtains
%\begin{widetext}
%%%%%%%%%%%%%%%%%%%%%%%%%%%%%%%%%%%%%%%
\begin{align}
    \frac{\partial\,\langle\whEu\rangle}{\partial t}\, =\, &
 \,\langle\wtvarrho\,\whuiv\cdot\whFextv\rangle\,
 +\,\langle\widehat{\jv}_\mathrm{i}\cdot\whEv\rangle\,
 +\,\langle\wtbPii:\grad\whuiv\rangle\nonumber\\
 &-\,\langle\widehat{\jv}_\mathrm{i}\cdot\bb{\epsilon}_\mathrm{MHD}^*\rangle\,
 +\,\langle\Tuu:\grad\whuiv\rangle\,,\label{eq:NHK_energy-summary-ionkin-filtered-avg}\\
  %%%%
    \frac{\partial\,\langle\whEPii\rangle}{\partial t}\, =\, & 
  \,-\langle\wtbPii:\grad\whuiv\rangle\,
  -\,\langle\TPiinablau\rangle\,,\label{eq:NHK_energy-summary-iontherm-filtered-avg}\\
  %%%%
  \frac{\partial\,\langle\whEue\rangle}{\partial t}\, =\, &
  \,\langle\widehat{\jv}_\mathrm{e}\cdot\whEv\rangle\,
  +\,\langle\wtbPie:\grad\whuev\rangle\nonumber\\
  &-\langle\widehat{\jv}_\mathrm{e}\cdot(\bb{\epsilon}_\mathrm{MHD}^*+\bb{\epsilon}_\mathrm{Hall}^*)\rangle
  +\varepsilon_m\langle\Tueue:\grad\whuev\rangle\,,
\label{eq:NHK_energy-summary-eleckin-filtered-avg}\\
  %%%%
    \frac{\partial\,\langle\whEPie\rangle}{\partial t}\, =\,& \left\{\begin{array}{l}
       (1-\gamma)\frac{3}{2}\Big[\langle\wtPe\big(\grad\cdot\whuev\big)\rangle
       \,+\langle\TPenablaue\rangle\Big]\\
       \,\\
      -\langle\wtbPie:\grad\whuev\rangle\,
      -\,\langle\TPienablaue\rangle
  \end{array}\right.\begin{array}{c}
        \quad\textrm{(a)}\\
        \, \\
        \quad\textrm{(b)}
  \end{array}\,,\label{eq:NHK_energy-summary-electherm-filtered-avg}\\
  %%%%
  \ \frac{\partial\,\langle\whEB\rangle}{\partial t}\,=\, & 
  \,-\,\langle\widehat{\jv}_\mathrm{i}\cdot\whEv\rangle\,
  -\,\langle\widehat{\jv}_\mathrm{e}\cdot\whEv\rangle\,
  -\,\langle\jv^*\cdot\whEv\rangle\label{eq:NHK_energy-summary-magn-filtered-avg}\,.
\end{align}
%%%%%%%%%%%%%%%%%%%%%%%%%%%%%%%%%%%%%%%
%\end{widetext}
In Figure \ref{fig:NHKenergychannels} we show a schematic view of the link between different energy channels in the NHK model as described by equations (\ref{eq:NHK_energy-summary-ionkin-filtered-avg})--(\ref{eq:NHK_energy-summary-magn-filtered-avg}), when the electron-fluid description includes finite-inertia effects, $\varepsilon_m\neq0$, and a complete pressure-tensor dynamics (case (b) in equation (\ref{eq:NHK_energy-summary-electherm-filtered-avg}); see also \citet{ChasapisAPJ2018} for a somewhat similar scheme, although not explicit in scale transfer).
In this regard, we remind the reader that similar analysis have previously obtained somewhat analogous fundamental equations and decomposition (see, e.g., \citet{YangPOP2017} and \citet{DelSartoPegoraroMNRAS2018}), although without specializing to the case of hybrid-kinetic models. In particular, by considering hybrid closures, here we have also explicitly addressed the effects associated to different electron closures (i.e., at the internal energy level) and those effects associated to the mass ratio, $\varepsilon_m=\me/\mi$ (i.e., on the non-ideal turbulent electric field that mediates the energy transfer through scales).

\begin{figure*}
    %\centering
    \includegraphics[width=\columnwidth]{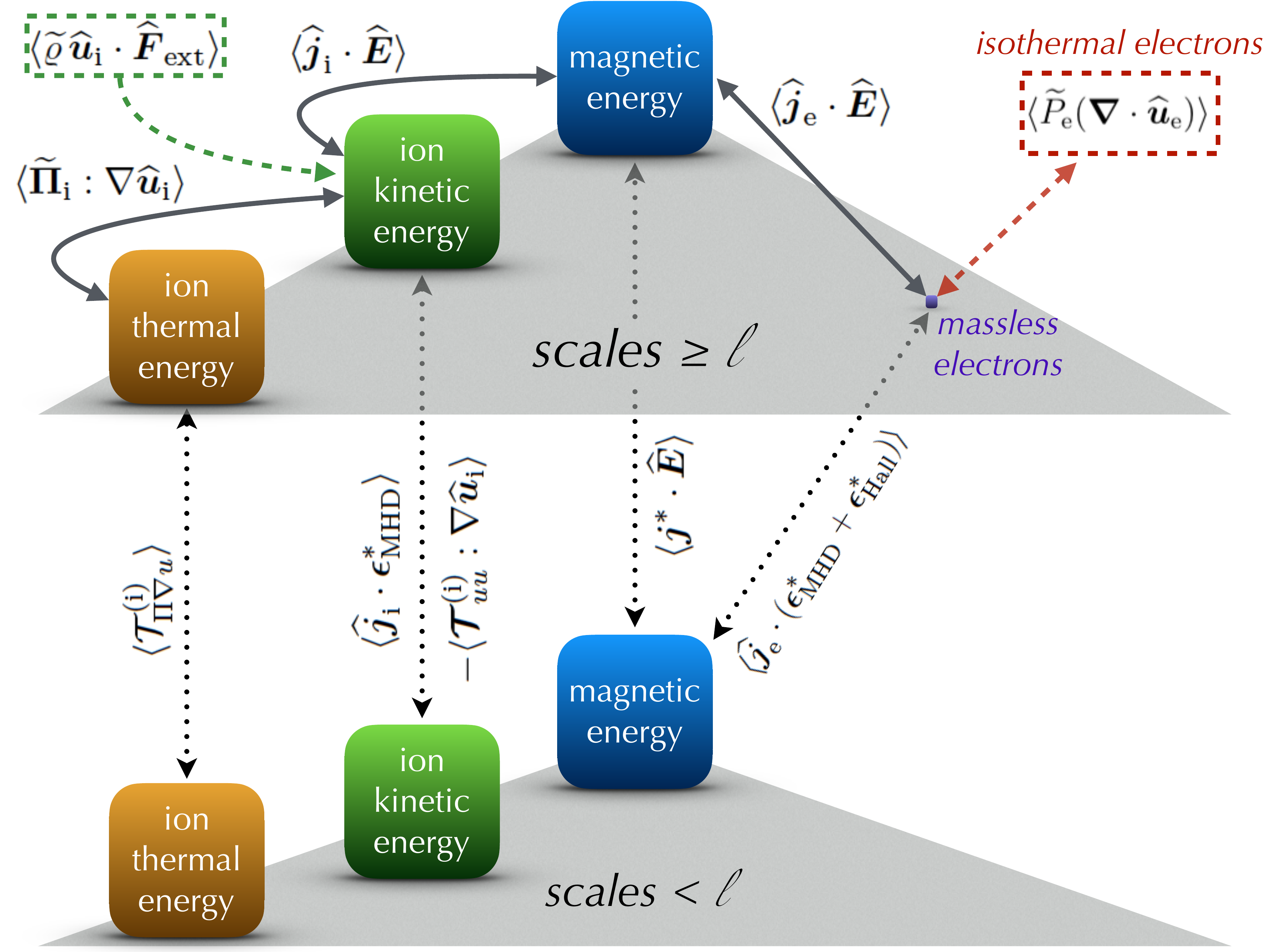}%
    \caption{Same Figure \ref{fig:NHKenergychannels} for the limiti of isothermal, massless electrons, $\gamma=1$ and $\varepsilon_m=0$ (see also Appendix~\ref{app:sec:HK-resistive}). This formalism shows that, in this limit, the turbulent energy transfer across scales is largely due to its additive incompressive contributions from the MHD and Hall transfer.}
    \label{fig:NHKenergychannels_masslessisothermal}
\end{figure*}

In this context, it is helpful to provide an explicit mention to the limit of isothermal, massless electrons, $\gamma=1$ and $\varepsilon_m=0$ (see also Appendix~\ref{app:sec:HK-resistive}). In this limit, the electron-thermal energy channel does not evolve, $\partial\langle\whEPie\rangle/\partial t=0$ (i.e. the electron thermal energy is kept constant, which in turn formally requires the appearance of a source term, $\langle\widehat{\cal I}_{\rm e}\rangle=\langle\widetilde{P}_{\rm e}(\grad\cdot\whuev)\rangle$).
Simultaneously, the electron-kinetic energy also goes to zero in this limit. As a result, equation (\ref{eq:NHK_energy-summary-eleckin-filtered-avg}) effectively becomes a sort of ``vertex-conservation law'' through the ``zero-measure'' electron-kinetic energy channel, $\langle\widehat{\jv}_\mathrm{e}\cdot\whEv\rangle+\langle\widetilde{P}_{\rm e}(\grad\cdot\whuev)\rangle-\langle\widehat{\jv}_\mathrm{e}\cdot(\bb{\epsilon}_\mathrm{MHD}^*+\bb{\epsilon}_\mathrm{Hall}^*)\rangle=0$, that instantaneously affects the magnetic-energy channel:
%%%%%%%%%%%%%%%%%%%%%%%%%%%%%%%%%%%%%%%
\begin{equation}\label{eq:NHK_energy-summary-magn-filtered-avg-masslessisothermal}
  \frac{\partial\,\langle\whEB\rangle}{\partial t}\bigg|_{\varepsilon_m=0,\gamma=1}\,=\, 
  \langle\widetilde{P}_{\rm e}(\grad\cdot\whuev)\rangle\,
  -\,\langle\widehat{\jv}_\mathrm{i}\cdot\whEv\rangle\,
  -\,\langle\widehat{\jv}_\mathrm{e}\cdot(\bb{\epsilon}_\mathrm{MHD}^*+\bb{\epsilon}_\mathrm{Hall}^*)\rangle\,
  -\,\langle\jv^*\cdot\whEv\rangle\,.
\end{equation}
%%%%%%%%%%%%%%%%%%%%%%%%%%%%%%%%%%%%%%%
A schematic view of the energy channels and transfer through scales in this limit is provided in Figure~\ref{fig:NHKenergychannels_masslessisothermal}.
This formalism also provides an evidence that, in a hybrid model with isothermal, massless electrons, the turbulent energy transfer through scales could be approximated by its additive incompressive contributions due to the MHD and Hall transfer~\citep{GaltierPRE2008,AndresPRE2018,SahraouiRvMPP2020} (as, for instance, it was assumed by \citet{HellingerAPJL2018}).
In fact, interestingly enough, a similar difference in the sub-grid turbulent electric field (and thus in the associated scale-to-scale magnetic-energy transfer) can be obtained from the Hall-MHD equations. This in turn may explain why Hall-MHD and hybrid-kinetics with isothermal, massless electrons have shown to provide similar results in terms of the turbulent cascade of magnetic-field fluctuations across the so-called ion break (see, e.g., \citet{PapiniAPJ2019}). In this context, the relevance and nature of the Hall electric-field fluctuations in possibly playing a relevant role in the transition to a so-called reconnection-mediated regime~\citep{CerriCalifanoNJP2017,FranciAPJL2017,LoureiroBoldyrevAPJ2017,MalletJPP2017} of sub-ion-scale turbulent energy transfer was already pointed out by means of hybrid-kinetic simulations in \citet{CerriCalifanoNJP2017}).

\subsection{Localized scale-to-scale energy transfer and the direction of the cascade}

Because the treatment of the space-filtered technique has been kept general so far, the exact meaning of the quantities on the right-hand side of the filtered equations slightly depends on the particular choice of the filter function $G$ in equation (\ref{eq:filter-def-1}).
However, let us consider a simple case in which $G$ is a low-pass filter in Fourier space, with $\ell_*$ being the characteristic scale below which the quantities are filtered out.
In such case, the above equations not only tell us how the different energy channels are coupled on the scales $\ell\geq\ell_*$, but it also highlights how the energy is transferred across the scale $\ell_*$ through the sub-grid terms (i.e., if the energy flux goes towards smaller or larger scales).
By considering various scales $\ell_*$ within the system, one can obtain a map of how different terms contribute to the energy flux through scales at each scale. 
Moreover, since the non-averaged sub-grid terms responsible for such energy transfer keep their real-space dependence, one can also investigate how localized this transfer is, as, e.g., its correlation with current and vorticity sheets, or with other magnetic and flow structures.
This is a very useful piece of information, as it enable us to investigate the presence of direct and inverse cascades in kinetic plasma turbulence and/or, by adopting the appropriate filters, also the occurrence of cross-scale interactions (i.e., ``non-local'' in Fourier space).

\section{Conclusions}\label{sec:discussion-conclusion}

In this work, we have derived the space-filtered (or, ``coarse-grained'') equations for a general class of quasi-neutral hybrid-kinetic (NHK) models, including their explicit formulation for specific hybrid-kinetic (HK) sub-models that are often adopted for the study of kinetic turbulence in collisionless plasmas~\cite[e.g.,][]{ServidioJPP2015,FranciAPJ2015,CerriAPJL2017,ArzamasskiyAPJ2019}, as well as for the full-kinetic (FK) case (for which large 3D simulations have become recently possible~\cite[e.g.,][]{MunozBuchnerPRE2018,GroseljPRL2018,GroseljPRX2019}).
This theoretical framework indeed represents an extremely useful tool for investigating the mechanisms underlying the energy transfer in plasma turbulence. 
In fact, the space-filtered technique allows to explicitly separate the contribution of different ``energy channels'', to elucidate their localization and correlation with ``coherent structures'' in real-space, to highlight the presence of both inverse and direct cascades, as well as the behavior of the energy flux through scales (e.g., if the constant cascade rate is a well-posed assumption, or if energy dissipation is a multi-scale process rather than being localized at certain specific scales~\citep[e.g.,][]{CerriPOP2014a,ToldPRL2015,TeacaNJP2017,HowesJPP2018}).

Despite the intrinsic value of the space-filtered approach for turbulence studies, it has not received widespread attention by the plasma physics community: only recently, it has been considered in the context of extended-fluid and/or of kinetic turbulence~\citep{BanonNavarroPOP2014,YangPOP2017,CamporealePRL2018,EyinkPRX2018}. 
Because this increasing (although underrated) attention to the filtered approach, we believe that it is of interest to provide its explicit formulation for a wide range of quasi-neutral hybrid-kinetic models that are commonly adopted for the investigation of space and astrophysical plasma turbulence at kinetic scales (see, e.g., \citet{CerriFrASS2019} and references therein).
For instance, within this formalism we can indeed explain why Hall-MHD and hybrid-kinetics with isothermal, massless electrons seem to provide similar results in terms of the turbulent cascade of magnetic-field fluctuations (see, e.g., \citet{PapiniAPJ2019}) and why this transfer may be approximated by the additive contributions due to the MHD and Hall terms in the sub-grid electric field (see, e.g., \citet{HellingerAPJL2018}). 
Indeed, the relevance and nature of the Hall electric-field fluctuations in possibly mediating the sub-ion-scale turbulent energy transfer in hybrid-kinetic turbulence was already pointed out by \citet{CerriCalifanoNJP2017} in the context of a so-called reconnection-mediated regime~\citep{CerriCalifanoNJP2017,FranciAPJL2017,LoureiroBoldyrevAPJ2017,MalletJPP2017}.

Once again, we stress that the fundamental difference with a Large-Eddy Simulation (LES) approach is that when the filtering scale is well resolved within a simulation domain, the calculation of the sub-grid source terms in (\ref{eq:NHK_whcalSsg-0})--(\ref{eq:NHK_whcalSsg-me}) becomes straightforward. The analysis of such terms in configuration space provides valuable information on how and where the energy cascade and coupling takes place.

\acknowledgements
S.S.C. was supported by NASA Grant No.~NNX16AK09G issued through the Heliophysics Supporting Research Program and by Max-Planck/Princeton Center for Plasma Physics (NSF grant PHY-1804048). 
E.C. is partially funded by NWO-Vidi grant 639.072.716.

\section*{DATA AVAILABILITY}
Data sharing is not applicable to this article as no new data were created or analyzed in this study

\appendix

\section{Explicit formulation for the hybrid-Vlasov-Maxwell (HVM) model in \citet{ValentiniJCP2007}}\label{app:sec:HVM}

The HVM model equations of \citet{ValentiniJCP2007}, inlcuding the external forcing term given in \citet{CerriAPJL2016}, read as equations (\ref{eq:NHK_Vlasov})--(\ref{eq:NHK_Maxwell}), but with the following approximated generalized Ohm's law:
%%%%%%%%%%%%%%%%%%%%%%%%%%%%%%%%%%%%%%%
\begin{equation}\label{app:eq:HVM_Ohm}
(1-d_{\rm e,0}^2\nabla^2)\Ev\,=\,
-\frac{\uev}{c}\times\Bv
-\frac{\grad P_{\rm e}}{en}+\frac{\varepsilon_m}{en}\grad\cdot\big(\bb{D}_J+\bPii\big)\,,
\end{equation}
%%%%%%%%%%%%%%%%%%%%%%%%%%%%%%%%%%%%%%%
where $d_{\rm e,0}^2=\epsilon_m d_{\rm i,0}^2=\varepsilon_m\mi c^2/4\pi e^2n_0$ is the electron inertial length computed with (homogeneous) background density, and we have introduced the tensor
\begin{equation}
 \bb{D}_J\equiv\varrho(\uiv\uiv-\uev\uev)=\frac{\mi}{e}\left(\Jv\uiv+\uiv\Jv-\frac{\Jv\Jv}{en}\right)
\end{equation}
for shortness.
This tensor is related to the (difference of the) Reynolds stress tensor of the ion and electron flows, and it can be seen as a Reynolds stress associated to the current-current and current-flow fluctuations' non-linearity.
The scalar electron pressure, $P_{\rm e}$, is closed via polytropic relation, as in equation (\ref{eq:NHK_Pe-polytropic}), and the $\nabla^2\Ev$ term can be obtained by using the identity
%%%%%%%%%%%%%%%%%%%%%%%%%%%%%%%%%%%%%%%
\begin{equation}\label{eq:HVM_nabla2E-term}
    \frac{\partial\,\Jv}{\partial t}\,=
    \frac{c}{4\pi}\grad\times\frac{\partial\,\Bv}{\partial t}\,=
    -\frac{c^2}{4\pi}\grad\times\grad\times\Ev
    \,,
    %\,=\,\frac{c^2}{4\pi}\Big(\nabla^2\Ev-\grad(\grad\cdot\Ev)\Big)\,,
\end{equation}
%%%%%%%%%%%%%%%%%%%%%%%%%%%%%%%%%%%%%%%
and then neglecting the $\grad\cdot\Ev$ by means of the quasi-neutrality approximation, $\partial_t\Jv\approx c^2\nabla^2\Ev/4\pi$.
Note that this version of the generalized Ohm's law can be obtained from (\ref{eq:NHK_Ohm_2}) with the approximation $(1+\varepsilon_m)^{-1}\simeq1$ and by neglecting inhomogeneities in front of the Laplacian term, i.e. $d_{\rm e}=\varepsilon_m\mi c^2/4\pi e^2n\simeq\varepsilon_m\mi c^2/4\pi e^2n_0=d_{\rm e,0}$. 

By following the same procedure as in Section \ref{subsec:NHK_EnergyEqs}, one easily derives the energy equations for the HVM model, given the generalized Ohm's law in (\ref{app:eq:HVM_Ohm}):
%%%%%%%%%%%%%%%%%%%%%%%%%%%%%%%%%%%%%%%
\begin{align}
%%%%%
  \frac{\partial\,{\cal E}_{u_{\rm i}}}{\partial t}\, 
  +\, \grad\cdot\big({\cal E}_{u_{\rm i}}\uiv+\bPii\cdot\uiv\big)\, 
  =\, & \,\varrho\uiv\cdot\Fextv+\bb{j}_{\rm i}\cdot\Ev
  \,+\,\bPii:\grad\uiv\,,\label{app:eq:HVM_E-ui}\\
%%%%%
 \frac{\partial\,{\cal E}_{\Pi_{\rm i}}}{\partial t}\, 
  +\, \grad\cdot\big({\cal E}_{\Pi_{\rm i}}\bb{u}_{\rm i}+\bb{q}_{\rm i}\big)\, 
  =\, & \,-\bb{\Pi}_{\rm i}:\grad\bb{u}_{\rm i}\,,\label{app:eq:HVM_E-Pi}\\
%%%%%
  \frac{\partial\,{\cal E}_{u_{\rm e}}}{\partial t} 
  + \grad\cdot\Big[\big({\cal E}_{u_{\rm e}}+P_{\rm e}\big)\uev\Big]\, 
  =\, & \,\varepsilon_m\varrho\uev\cdot\Fextv+\,\bb{j}_{\rm e}\cdot\Ev\,+P_{\rm e}\grad\cdot\uev\nonumber\\
   & \, -\varepsilon_m\bb{j}_{\rm e}\cdot\Ev
  \,+\varepsilon_m\frac{\delta n}{n}\bb{j}_{\rm e}\cdot d_{\rm i,0}^2\nabla^2\Ev\,,\label{app:eq:HVM_E-ue}\\
%%%%%
  \frac{\partial\,{\cal E}_{P_{\rm e}}}{\partial t}\, 
  +\, \grad\cdot\big({\cal E}_{P_{\rm e}}\bb{u}_{\rm e}\big)\, 
  =\,& \,\big(1-\gamma\big){\cal E}_{P_{\rm e}}\grad\cdot\bb{u}_{\rm e}\,,\label{app:eq:HVM_E-Pe}\\
%%%%%
  \frac{\partial\,{\cal E}_B}{\partial t}\, 
  +\, \grad\cdot\left(\frac{\bb{E}\times\bb{B}}{4\pi}c\right)\, 
  =\, & \,-\,\bb{j}_{\rm e}\cdot\bb{E}\,-\,\bb{j}_{\rm i}\cdot\bb{E}\label{app:eq:HVM_E-B}\,,
\end{align}
%%%%%%%%%%%%%%%%%%%%%%%%%%%%%%%%%%%%%%%
where we have introduced the species' current density, $\bb{j}_\alpha\equiv e_\alpha n\uav$ and the density fluctuations, $\delta n\equiv n-n_0$.
Note that the last term in equation (\ref{app:eq:HVM_E-ue}) for the electron kinetic energy density is the result of the approximations made in (\ref{app:eq:HVM_Ohm}) with respect to its complete version in (\ref{eq:NHK_Ohm_2}).
Namely, the extra $-\varepsilon_m\bb{j}_e\cdot\Ev$ is a consequence of the approximation $(1+\varepsilon_m)^{-1}\simeq1$, while the term proportional to $\delta n/n$ comes from the approximation $d_{\rm e}\simeq d_{\rm e,0}$ (i.e., $1/n\simeq1/n_0$) in the Laplacian term deriving from $\partial\Jv/\partial t$.

The equation for the total energy density in HVM then reads:
\begin{widetext}
%%%%%%%%%%%%%%%%%%%%%%%%%%%%%%%%%%%%%%%
\begin{equation}\label{app:eq:HVM_TotalEn}
  \frac{\partial\,{\cal E}}{\partial t}\, 
  +\, \grad\cdot\bb{\Psi}_{\cal E}\,=\,
  \varrho\big(\uiv+\varepsilon_m\uev\big)\cdot\Fextv\, +\,\left(\frac{5}{3}-\gamma\right){\cal E}_{P_{\rm e}}\big(\grad\cdot\bb{u}_{\rm e}\big)\,
  -\,\varepsilon_m\bb{j}_{\rm e}\cdot\left(1-\frac{\delta n}{n}d_{\rm i,0}^2\nabla^2\right)\Ev\,,
\end{equation}
%%%%%%%%%%%%%%%%%%%%%%%%%%%%%%%%%%%%%%%
\end{widetext}
where the total energy density flux, $\bb{\Psi}_{\cal E}$, is
%%%%%%%%%%%%%%%%%%%%%%%%%%%%%%%%%%%%%%%
\begin{equation}\label{app:eq:HVM_TotalEnFlux}
  \bb{\Psi}_{\cal E}\, 
  =\,\big({\cal E}_{u_{\rm i}}+{\cal E}_{\Pi_{\rm i}}\big)\uiv
  +\bPii\cdot\uiv
  +\qv_{\rm i}\,+\frac{5}{3}{\cal E}_{P_{\rm e}}\uev
  +\frac{\bb{E}\times\bb{B}}{4\pi}c\,,
\end{equation}
%%%%%%%%%%%%%%%%%%%%%%%%%%%%%%%%%%%%%%%

\subsection{Space-filtered equations for HVM}\label{app:subsec:HVM_EnEqs_SpaceFiltered}

Before presenting the space-filtered version of the energy equations, it is worth making some explicit considerations on the filtered electric field from the generalized Ohm's law adopted by the HVM model.

\subsubsection{The sub-grid (``turbulent'') electric field in HVM}

By multiplying the generalized Ohm's law in (\ref{app:eq:HVM_Ohm}) and applying the filter to it, one obtain its filtered version, i.e.,
%%%%%%%%%%%%%%%%%%%%%%%%%%%%%%%%%%%%%%%
\begin{equation}\label{app:eq:HVM_Ohm-filtered}
 \big(1-\varepsilon_m d_{\rm i,0}\nabla^2\big)\whEv\,=\, -\,\frac{\whuev\times\whBv}{c}\,
 -\,\frac{\mi}{e}\frac{\grad\wtPe}{\wtvarrho}\,
 +\,\varepsilon_m\frac{\mi}{e}\frac{1}{\wtvarrho}\grad\cdot\big(\widehat{\bb{D}}_J+\wtbPii\big)\,+\,\bb{\epsilon}^*\,.
\end{equation}
%%%%%%%%%%%%%%%%%%%%%%%%%%%%%%%%%%%%%%%
where we have defined $\widehat{\bb{D}}_J=\wtvarrho(\whuiv\whuiv-\whuev\whuev)$ and, analogously to what was done in Section~\ref{subsec:NHK_E-field-filtered}, the sub-grid (``turbulent'') electric field as $\bb{\epsilon}^*=\bb{\epsilon}_{\mathrm{MHD}}^*+\bb{\epsilon}_{\mathrm{Hall}}^*+,\bb{\epsilon}_{d_\mathrm{e}}^*$ with
%%%%%%%%%%%%%%%%%%%%%%%%%%%%%%%%%%%%%%%
\begin{align}
  \bb{\epsilon}_{\mathrm{MHD}}^*\,=\, & -\TuxB\label{app:eq:HVM_Emhd-subgrid}\\
  \bb{\epsilon}_{\mathrm{Hall}}^*\,=\, & -\TJxB\label{app:eq:HVM_Ehall-subgrid}\\
  \bb{\epsilon}_{d_\mathrm{e}}^*\,=\, & -\varepsilon_m\frac{\mi}{e}\frac{1}{\wtvarrho}\grad\cdot\big(\Tueue-\Tuu\big)\,=\,-\varepsilon_m\frac{\mi}{e}\frac{1}{\wtvarrho}\grad\cdot\big(\TJJ-\TJuuJ\big)\label{app:eq:HVM_Ede-subgrid}\,.
\end{align}
%%%%%%%%%%%%%%%%%%%%%%%%%%%%%%%%%%%%%%%
Note that, with respect to (\ref{eq:NHK_Ede-subgrid-2}), in HVM the electron-inertia contribution $\bb{\epsilon}_{d_\mathrm{e}}$ effectively misses the contribution from the turbulent (sub-grid) Reynolds stress associated to the ion-flow fluctuations.

\subsubsection{The filtered total energy equation for HVM}

We now follow the same procedure as in Section \ref{subsec:NHK_EnEqs_SpaceFiltered} to derive the filtered version of the energy equations for the HVM model:
%%%%%%%%%%%%%%%%%%%%%%%%%%%%%%%%%%%%%%%
\begin{align}
 \frac{\partial\,\whEu}{\partial t}\, +\, 
  \grad\cdot\Big[\whEu\whuiv+\big(\wtbPii+\Tuu\big)\cdot\whuiv\Big]\, =\, 
  &\,\wtvarrho\whuiv\cdot\whFextv
   +\,\wtbPii:\grad\whuiv\,
   +\widehat{\bb{j}}_{\rm i}\cdot\whEv\nonumber\\
 & \, +\,\Tuu:\grad\whuiv\,
 -\,\widehat{\bb{j}}_{\rm i}\cdot\bb{\epsilon}_{\rm MHD}^*\,,\label{app:eq:HVM_E-ui_filtered}\\
%%%%%
 \frac{\partial\,\whEPii}{\partial t}\, +\, 
  \grad\cdot\Big(\widehat{{\cal E}_{\Pi_{\rm i}}\uiv}+\whqiv\Big)\, =\, & 
  \,-\wtbPii:\grad\whuiv\, -\,
  \TPiinablau\,,\label{app:eq:HVM_E-Pi_filtered}\\
%%%%
 \frac{\partial\,\whEue}{\partial t}+ 
  \grad\cdot\Big[\big(\whEue+\wtPe\big)\whuev+\varepsilon_m\Tueue\cdot\whuev\Big] = &
  \,\widehat{\bb{j}}_{\rm e}\cdot\whEv\,
  +\,\wtPe\grad\cdot\whuev\,
  +\,\varepsilon_m\wtvarrho\whuev\cdot\whFextv\nonumber\\
 & -\widehat{\bb{j}}_{\rm e}\cdot\big(\bb{\epsilon}_{\rm MHD}^*+\bb{\epsilon}_{\rm Hall}^*\big)\,
 +\varepsilon_m\Tueue:\grad\whuev\nonumber\\
 &-\varepsilon_m\widehat{\bb{j}}_{\rm e}\cdot\left[\left(1-\frac{\widetilde{\delta\varrho}}{\wtvarrho}d_{\rm i,0}\nabla^2\right)\whEv-\bb{\epsilon}_{\rm MHD}^*-\bb{\epsilon}_{\rm Hall}^*-\bb{\epsilon}_{\nabla^2}^*\right]\,,\label{app:eq:HVM_E-ue_filtered}\\
%%%%%
    \frac{\partial\,\whEPe}{\partial t}\, +\,
  \grad\cdot\Big(\widehat{{\cal E}_{P_{\rm e}}\uev}\Big)\, =\, &
  \,(1-\gamma)\Big[\wtEPe\big(\nabla\cdot\whuev\big)
  +\TPenablaue\Big]\,,\label{app:eq:HVM_E-Pe_filtered}\\
\frac{\partial\,\whEB}{\partial t}\, 
  +\, \grad\cdot\left(\frac{\whEv\times\whBv}{4\pi}c\right)\, 
  =\,& -\widehat{\bb{j}}_{\rm i}\cdot\whEv-\widehat{\bb{j}}_{\rm e}\cdot\whEv-\bb{j}^*\cdot\whEv\label{app:eq:HVM_E-B_filtered}\,.
\end{align}
%%%%%%%%%%%%%%%%%%%%%%%%%%%%%%%%%%%%%%%
where $\widehat{\bb{j}}_\mathrm{i}=\frac{e}{\mi}\wtvarrho\,\whuiv$ and we have introduced $\widetilde{\delta\varrho}=\wtvarrho-\varrho_0$ and an equivalent sub-grid electric field that arises from the Laplacian term in the filtered electron-momentum equation (again, related to the homogeneity approximation on the $d_{\rm e}^2\nabla^2\Ev$ term in (\ref{app:eq:HVM_Ohm})), i.e.,
%%%%%%%%%%%%%%%%%%%%%%%%%%%%%%%%%%%%%%%
\begin{equation}\label{app:eq:HVM_Enabla2-subgrid}
  \bb{\epsilon}_{\nabla^2}^*\,=\, 
  \frac{\varrho_0}{\wtvarrho}d_{\rm i,0}^2\nabla^2\,\bb{\cal T}_{E}\,.
\end{equation}
%%%%%%%%%%%%%%%%%%%%%%%%%%%%%%%%%%%%%%%
with $\bb{\cal T}_{E}\equiv(\widetilde{\varrho\Ev}-\wtvarrho\widetilde{\Ev})/\wtvarrho=\whEv-\widetilde{\Ev}$.
Summing up equations (\ref{app:eq:HVM_E-ui_filtered})--(\ref{app:eq:HVM_E-B_filtered}), one obtains the filtered version of the total energy density equation:
%%%%%%%%%%%%%%%%%%%%%%%%%%%%%%%%%%%%%%%
\begin{equation}\label{app:eq:HVM_TotalEn-filtered}
  \frac{\partial\,\widehat{\cal E}}{\partial t}\, 
  +\, \grad\cdot\widehat{\bb{\Psi}}_{\widehat{\cal E}}\, =\,
  \wtvarrho\,\big(\whuiv+\varepsilon_m\whuev\big)\cdot\whFextv\,+
  \,\left(\frac{5}{3}-\gamma\right)\wtEPe\big(\grad\cdot\whuev\big)\,
  +\,\whcalSsg[0]\,
  +\,\whcalSsg[m_{\rm e}]\,
  +\,\whcalSsg[{\rm HVM}]\,,
\end{equation}
%%%%%%%%%%%%%%%%%%%%%%%%%%%%%%%%%%%%%%%
with $\widehat{\cal E}=\whEu+\whEPii+\whEue+\whEPe+\whEB$ and
%%%%%%%%%%%%%%%%%%%%%%%%%%%%%%%%%%%%%%%
\begin{equation}\label{eq:HVM_TotalEnFlux-filtered}
  \widehat{\bb{\Psi}}_{\widehat{\cal E}}\,=\,
  \,\whEu\whuiv\,
  +\,\big(\wtbPii+\Tuu\big)\cdot\whuiv\,
  +\,\widehat{{\cal E}_{\Pi_{\rm i}}\uiv}\,
  +\,\whqiv\,
  +\,\frac{5}{3}\whEue\whuev\,
  +\,\varepsilon_m\Tueue\cdot\whuev\,
  +\,\widehat{{\cal E}_{P_{\rm e}}\uev}\,
  +\,\frac{\whEv\times\whBv}{4\pi}c\,,
\end{equation}
%%%%%%%%%%%%%%%%%%%%%%%%%%%%%%%%%%%%%%%
while the sub-grid terms are given by
%%%%%%%%%%%%%%%%%%%%%%%%%%%%%%%%%%%%%%%
\begin{equation}
  \whcalSsg[0]=
  -(\widehat{\bb{j}}_{\rm i}+\widehat{\bb{j}}_{\rm e})\cdot\bb{\epsilon}_{\rm MHD}^*\,
  -\,\widehat{\bb{j}}_{\rm e}\cdot\bb{\epsilon}_{\rm Hall}^*\,
  -\,\bb{j}^*\cdot\whEv\,
  +\,\Tuu:\grad\whuiv\,
  -\,\TPiinablau\,
  +\,(1-\gamma)\TPenablaue\,,\label{app:eq:HVM_whcalSsg-0}
\end{equation}
%%%%%%%%%%%%%%%%%%%%%%%%%%%%%%%%%%%%%%%
%%%%%%%%%%%%%%%%%%%%%%%%%%%%%%%%%%%%%%%
\begin{equation}\label{app:eq:HVM_whcalSsg-me}
  \whcalSsg[m_{\rm e}]\, =\,
  \varepsilon_m\Tueue:\grad\whuev\,
\end{equation}
%%%%%%%%%%%%%%%%%%%%%%%%%%%%%%%%%%%%%%%
%%%%%%%%%%%%%%%%%%%%%%%%%%%%%%%%%%%%%%%
\begin{equation}
  \whcalSsg[{\rm HVM}]=
  -\varepsilon_m\widehat{\bb{j}}_{\rm e}\cdot\left[\left(1-\frac{\widetilde{\delta\varrho}}{\wtvarrho}d_{\rm i,0}\nabla^2\right)\whEv-\bb{\epsilon}_{\rm MHD}^*-\bb{\epsilon}_{\rm Hall}^*-\bb{\epsilon}_{\nabla^2}^*\right]\,,\label{app:eq:HVM_whcalSsg-HVM}
\end{equation}
%%%%%%%%%%%%%%%%%%%%%%%%%%%%%%%%%%%%%%%
where the first two sub-grid terms, $\whcalSsg[0]$ and $\whcalSsg[m_{\rm e}]$, are indeed the same as the ones in (\ref{eq:NHK_whcalSsg-0})--(\ref{eq:NHK_whcalSsg-me}) for a generic NHK model, while the second term, $\whcalSsg[{\rm HVM}]$, is a model-dependent term, i.e., specific of the HVM model.
Also note that this additional sub-grid term is strictly related to finite-inertia effects in the HVM's electron fluid and thus it disappears in the massless electrons limit, $\varepsilon_m\to0$.

\section{Formulation for massless electrons and resisitive Ohm's law}\label{app:sec:HK-resistive}

It is useful to provide the space-filtered equations also in the limit of massless electrons and a generalized Ohm's law that includes a resistive term:
%%%%%%%%%%%%%%%%%%%%%%%%%%%%%%%%%%%%%%%
\begin{equation}\label{app:eq:Camelia-Pegasus_Ohm}
\Ev\,=\,
\,-\frac{\uiv}{c}\times\Bv\,
+\,\frac{\Jv\times\Bv}{enc}\,
-\,\frac{\grad P_{\rm e}}{en}\,
+\,\eta\Jv\,,
\end{equation}
%%%%%%%%%%%%%%%%%%%%%%%%%%%%%%%%%%%%%%%
and a polytropic closure is adopted for the scalar electron pressure.
This version of the generalized Ohm's law is indeed widely adopted in numerical implementations of hybrid-kinetic models~\citep[e.g.,][]{KunzJCP2014,ServidioJPP2015,FranciAPJ2015}. 

In this case, the total energy equation reads
%%%%%%%%%%%%%%%%%%%%%%%%%%%%%%%%%%%%%%%
\begin{align}\label{app:eq:Camelia-Pegasus_TotalEn}
  \frac{\partial\,{\cal E}}{\partial t}\, 
  +\, \grad\cdot\bb{\Psi}_{\cal E}\, 
  =\, &
  \,\varrho\bb{u}_{\rm i}\cdot\bb{F}_{\rm ext}\,+\,
  \left(\frac{5}{3}-\gamma\right){\cal E}_{P_{\rm e}}\big(\grad\cdot\bb{u}_{\rm e}\big)\,-\,
  \eta\,\bb{j}_{\rm e}\cdot\Jv\nonumber\\
  =\, &
  \varrho\uiv\cdot\Fextv\,
  +\,\left(\frac{5}{3}-\gamma\right){\cal E}_{P_{\rm e}}\big(\grad\cdot\bb{u}_{\rm e}\big)\,
  +\,\eta\bb{j}_{\rm i}\cdot\Jv\,
  -\,\eta|\Jv|^2\,,
\end{align}
%%%%%%%%%%%%%%%%%%%%%%%%%%%%%%%%%%%%%%%
with
%%%%%%%%%%%%%%%%%%%%%%%%%%%%%%%%%%%%%%%
\begin{equation}\label{app:eq:Camelia-Pegasus_TotalEnFlux}
  \widehat{\bb{\Psi}}_{\widehat{\cal E}}\,=\,
  \,\big({\cal E}_{u_{\rm i}}+{\cal E}_{\Pi_{\rm i}}\big)\uiv\,
  +\,\bPii\cdot\uiv\,
  +\,\bb{q}_{\rm i}\,
  +\,\frac{5}{3}{\cal E}_{P_{\rm e}}\uev\,
  +\,\frac{\bb{E}\times\bb{B}}{4\pi}c\,.
\end{equation}
%%%%%%%%%%%%%%%%%%%%%%%%%%%%%%%%%%%%%%%
Note that, since $\jv_{\rm e}\cdot\Jv$ is not positive definite, the dissipative nature of the resistive term in (\ref{app:eq:Camelia-Pegasus_TotalEn}) is not mathematically ensured.
The space-filtered version of the equation for the total energy density above is
%%%%%%%%%%%%%%%%%%%%%%%%%%%%%%%%%%%%%%%
\begin{equation}\label{app:eq:Camelia-Pegasus_TotalEn-filtered}
  \frac{\partial\,\widehat{\cal E}}{\partial t}\, 
  +\, \grad\cdot\widehat{\bb{\Psi}}_{\widehat{\cal E}}\,
  =\,
  \wtvarrho\,\whuiv\cdot\whFextv\,
  +\,\left(\frac{5}{3}-\gamma\right)\wtEPe\big(\grad\cdot\whuev\big)\,
  -\,\eta\left|\whJv\right|^2\,
  +\,\eta\,\widehat{\jv}_{\rm i}\cdot\whJv\,
  +\,\whcalSsg[0]\,+\,\whcalSsg[\eta]\,,
\end{equation}
%%%%%%%%%%%%%%%%%%%%%%%%%%%%%%%%%%%%%%%
where the sub-grid term $\whcalSsg[0]$ corresponds to the one in (\ref{app:eq:HVM_whcalSsg-0}), while the resistive contribution to the sub-grid term, $\whcalSsg[\eta]$, is given by
%%%%%%%%%%%%%%%%%%%%%%%%%%%%%%%%%%%%%%%
\begin{equation}\label{app:eq:Camelia-Pegasus_whcalSsg-eta}
 \whcalSsg[\eta]\,=\,\eta\,\jv^*\cdot\whJv\,,
\end{equation}
%%%%%%%%%%%%%%%%%%%%%%%%%%%%%%%%%%%%%%%
where we remind that $\jv^*=\whJv-\widehat{\jv}_{\rm e}-\widehat{\jv}_{\rm i}$ is the sub-grid (``turbulent'') current density.  

\subsection{The Hall-MHD limit}

Another approximation often employed in hybrid-kinetic simulations consists of adopting the Hall-MHD limit of the generalized Ohm's law~\citep[see, e.g.,][and references therein]{PalmrothLRCA2018}:
%%%%%%%%%%%%%%%%%%%%%%%%%%%%%%%%%%%%%%%
\begin{equation}\label{app:eq:Vlasiator_Ohm}
\Ev\,=\,
\,-\frac{\uiv}{c}\times\Bv\,
+\,\frac{\Jv\times\Bv}{enc}\,
=\,-\frac{\uev}{c}\times\Bv\,,
\end{equation}
%%%%%%%%%%%%%%%%%%%%%%%%%%%%%%%%%%%%%%%
i.e., the limit of infinite conductivity and cold, massless electron fluid.
In this case, all the terms containing the electron pressure or the resistivity in the equations (\ref{app:eq:Camelia-Pegasus_TotalEn})--(\ref{app:eq:Camelia-Pegasus_TotalEnFlux}) vanish, and we obtain a simple total energy equation:
%%%%%%%%%%%%%%%%%%%%%%%%%%%%%%%%%%%%%%%
\begin{equation}\label{app:eq:Vlasiator_TotalEn}
  \frac{\partial\,{\cal E}}{\partial t}\, 
  +\, \grad\cdot\left[\big({\cal E}_{u_{\rm i}}+{\cal E}_{\Pi_{\rm i}}\big)\uiv
  +\bPii\cdot\bb{u}_{\rm i}
  +\bb{q}_{\rm i}
  +\frac{\bb{E}\times\bb{B}}{4\pi}c\right]\, 
  =\,
  \varrho\bb{u}_{\rm i}\cdot\bb{F}_{\rm ext}\,.
\end{equation}
%%%%%%%%%%%%%%%%%%%%%%%%%%%%%%%%%%%%%%%
Its filtered counterpart therefore reads
%%%%%%%%%%%%%%%%%%%%%%%%%%%%%%%%%%%%%%%
\begin{equation}\label{app:eq:Vlasiator_TotalEn-filtered}
  \frac{\partial\,\widehat{\cal E}}{\partial t}\, 
  +\, \grad\cdot\widehat{\bb{\Psi}}_{\widehat{\cal E}}\,
  =\,
  \wtvarrho\,\whuiv\cdot\whFextv\,
  +\,\whcalSsg[{\rm H-MHD}]\,,
\end{equation}
%%%%%%%%%%%%%%%%%%%%%%%%%%%%%%%%%%%%%%%
where the flux and the sub-grid term are give by
%%%%%%%%%%%%%%%%%%%%%%%%%%%%%%%%%%%%%%%
\begin{equation}\label{eq:Vlasiator_TotalEnFlux-filtered}
  \widehat{\bb{\Psi}}_{\widehat{\cal E}}\,=\,
  \whEu\whuiv\,
  +\,\widehat{{\cal E}_{u_{\rm i}}\uiv}\,
  +\,(\wtbPii+\Tuu)\cdot\whuiv\,
  +\,\whqiv\,
  +\,\frac{\whEv\times\whBv}{4\pi}c\,,
\end{equation}
%%%%%%%%%%%%%%%%%%%%%%%%%%%%%%%%%%%%%%%
and 
%%%%%%%%%%%%%%%%%%%%%%%%%%%%%%%%%%%%%%%
\begin{equation}\label{app:eq:Vlasiator_whcalSsg0-def}
  \whcalSsg[{\rm H-MHD}]=
  -(\widehat{\bb{j}}_{\rm i}+\widehat{\bb{j}}_{\rm e})\cdot\bb{\epsilon}_{\rm MHD}^*\,
  -\,\widehat{\bb{j}}_{\rm e}\cdot\bb{\epsilon}_{\rm Hall}^*\,
  -\,\bb{j}^*\cdot\whEv\,
  +\,\Tuu:\grad\whuiv\,
  -\,\TPiinablau\,.
\end{equation}
%%%%%%%%%%%%%%%%%%%%%%%%%%%%%%%%%%%%%%%

\section{Formulation for the full-kinetic case}\label{app:sec:fullkinetic}

For the sake of completeness, we also report here the space-filtered equations for a full-Vlasov plasma (see also \citet{YangPOP2017}).
When the full-kinetic case is considered (neglecting the external forcing for the sake of simplicity), the moments equations that have been derived from the Vlasov equation for the ions in Section~\ref{subsec:NHK_EnergyEqs} are now holding for each species $\alpha$ with mass $\ma$ and charge $e_\alpha$, i.e.
%%%%%%%%%%%%%%%%%%%%%%%%%%%%%%%%%%%%%%%
\begin{equation}\label{app:eq:FK_continuity}
\frac{\partial\,\varrho_\alpha}{\partial t}\, +\, \grad\cdot\big(\varrho_\alpha\uav\big)\, =\, 0\,,
\end{equation}
%%%%%%%%%%%%%%%%%%%%%%%%%%%%%%%%%%%%%%%
%%%%%%%%%%%%%%%%%%%%%%%%%%%%%%%%%%%%%%%
\begin{equation}\label{app:eq:FK_momentum}
  \frac{\partial\,(\varrho_\alpha\uav)}{\partial t}\, +\, \grad\cdot\big(\varrho_\alpha\uav\uav+\bPia\big)\, =\,
  \frac{e_\alpha}{\ma}\varrho_\alpha\left(\Ev+\frac{\uav}{c}\times\Bv\right)\,,
\end{equation}
%%%%%%%%%%%%%%%%%%%%%%%%%%%%%%%%%%%%%%%
%%%%%%%%%%%%%%%%%%%%%%%%%%%%%%%%%%%%%%%
\begin{equation}\label{app:eq:FK_PressTens}
  \frac{\partial\,\bPia}{\partial t}\, 
  +\, \grad\cdot\big(\bPia\uav+\bQa\big)\, 
  +\, \Big\{\big(\bPia\cdot\grad\big)\uav\Big\}^{\rm sym}=\,
  \Omegaca\,\big\{\bPia\times\bv\big\}^{\rm sym}\,,
\end{equation}
%%%%%%%%%%%%%%%%%%%%%%%%%%%%%%%%%%%%%%%
where $\varrho_\alpha=\ma n_\alpha$ is the species' mass density, and $\Omegaca=e_\alpha B/\ma c$ its gyro-frequency. From these equations and from Maxwell's equations (now including displacement current in the Amp\'ere's law), the energy equations are readily derived:
%%%%%%%%%%%%%%%%%%%%%%%%%%%%%%%%%%%%%%%
\begin{align}
  \frac{\partial\,{\cal E}_{u_\alpha}}{\partial t}\, 
  +\, \grad\cdot\big({\cal E}_{u_\alpha}\uav+\bPia\cdot\uav\big)\, 
  =\, & \,+\bPia:\grad\uav\,
  +\,e_\alpha n_\alpha\uav\cdot\Ev\,,\label{app:eq:FK_energy-summary-kin}\\
  %%%%
  \frac{\partial\,{\cal E}_{\Pi_\alpha}}{\partial t}\, 
  +\, \grad\cdot\big({\cal E}_{\Pi_\alpha}\uav+\bb{q}_\alpha\big)\, 
  =\, & \,-\bPia:\grad\uav\,,\label{app:eq:FK_energy-summary-therm}\\
  %%%%
  \frac{\partial\,{\cal E}_\mathrm{em}}{\partial t}\, 
  +\, \grad\cdot\left(\frac{\bb{E}\times\bb{B}}{4\pi}c\right)\, 
  =\, & \,-\bigg(\sum_\alpha e_\alpha n_\alpha\uav\bigg)\cdot\bb{E}\label{app:eq:FK_energy-summary-em}\,,
\end{align}
%%%%%%%%%%%%%%%%%%%%%%%%%%%%%%%%%%%%%%%
where ${\cal E}_{u_\alpha}=\ma n_\alpha u_\alpha^2/2$, ${\cal E}_{\Pi_\alpha}=\mathrm{tr}[\bPia]/2$, and ${\cal E}_\mathrm{em}=(E^2+B^2)/8\pi$.
From the above, proceeding as in Section~\ref{subsec:NHK_EnEqs_SpaceFiltered}, the space-filtered equations for the full-kinetic case read
%%%%%%%%%%%%%%%%%%%%%%%%%%%%%%%%%%%%%%%
\begin{align}
    \frac{\partial\whEua}{\partial t} + 
  \grad\cdot\Big[\whEua\whuav+\big(\wtbPia+\Tuaua\big)\cdot\whuav\Big] = &
  \big(\wtbPia+\Tuaua\big):\grad\whuav+\widehat{\jv}_\alpha\cdot\big(\whEv+\TuaxB\big),\label{app:eq:FK_energy-summary-kin-filtered}\\
  %%%%
    \frac{\partial\,\whEPia}{\partial t} + 
  \grad\cdot\Big(\widehat{{\cal E}_{\Pi_\alpha}\uav}+\whqav\Big)=\, & 
  \,-\wtbPia:\grad\whuav\,
  -\,\TPianablaua\,,\label{app:eq:FK_energy-summary-therm-filtered}\\
  %%%%
  \frac{\partial\,\whEem}{\partial t} 
  + \grad\cdot\left(\frac{\whEv\times\whBv}{4\pi}c\right) 
  =\, & 
  \,-\,\sum_\alpha\bigg(\widehat{\jv}_\alpha\cdot\whEv\,+\,\TJEa\bigg)\,,\label{app:eq:FK_energy-summary-em-filtered}
\end{align}
%%%%%%%%%%%%%%%%%%%%%%%%%%%%%%%%%%%%%%%
where $\widehat{\bb{j}}_\alpha\equiv\frac{e_\alpha}{\ma}\wtvarrho_\alpha\,\whuav=e_\alpha\widetilde{n}_\alpha\whuav$ is the current density of the $\alpha$ species, and the the sub-grid terms ${\cal T}$ are defined as
%%%%%%%%%%%%%%%%%%%%%%%%%%%%%%%%%%%%%%%
\begin{equation}\label{app:eq:FK_Tuaua-def}
  \Tuaua\,=\,
  \wtvarrho_\alpha\big(\reallywidehat{\uav\uav}\,
  -\,\whuav\whuav\big)\,,
\end{equation}
%%%%%%%%%%%%%%%%%%%%%%%%%%%%%%%%%%%%%%%
%%%%%%%%%%%%%%%%%%%%%%%%%%%%%%%%%%%%%%%
\begin{equation}\label{app:eq:FK_TuxB-def}
  \TuaxB\,=\,\frac{1}{c}\Big(
  \reallywidehat{\uav\times\Bv}\,
  -\,\whuav\times\whBv\Big)\,.
\end{equation}
%%%%%%%%%%%%%%%%%%%%%%%%%%%%%%%%%%%%%%%
%%%%%%%%%%%%%%%%%%%%%%%%%%%%%%%%%%%%%%%
\begin{equation}\label{app:eq:FK_TPianablaua-def}
  \TPianablaua\,=\,
  \reallywidehat{\bPia:\grad\uav}\,
  -\,\wtbPia:\grad\whuav\,,
\end{equation}
%%%%%%%%%%%%%%%%%%%%%%%%%%%%%%%%%%%%%%%
%%%%%%%%%%%%%%%%%%%%%%%%%%%%%%%%%%%%%%%
\begin{equation}\label{app:eq:FK_TJE-def}
  \TJE\,=\,
  \,\bigg[\sum_\alpha\frac{e_\alpha}{\ma}\big(\widehat{\varrho_\alpha\uav}-\wtvarrho_\alpha\,\whuav\big)\bigg]\cdot\whEv\,
  =\,\sum_\alpha \jv_\alpha^*\cdot\whEv\,
  =\,\sum_\alpha\TJEa\,,
\end{equation}
%%%%%%%%%%%%%%%%%%%%%%%%%%%%%%%%%%%%%%%
with $\bb{j}^*=\sum_\alpha\bb{j}_\alpha^*$ the sub-grid (or ``turbulent'') current density, as defined in (\ref{eq:NHK_J-subgrid}).

\section{Generalized Ohm's law in the quasi-neutral limit: equivalence between (\ref{eq:NHK_Ohm}) and (\ref{eq:NHK_Ohm_2})}\label{app:ohm}

We want to explicitly show that the two forms of the generalized Ohm's law, Eq.~(\ref{eq:NHK_Ohm}) and Eq.~(\ref{eq:NHK_Ohm_2}), are equivalent. 
In order to do this, let us first consider the ``classic'' derivation of the generalized Ohm's law from the two-fluid momentum equations~\citep[e.g.,][]{KrallTrivelpiece1973},
%%%%%%%%%%%%%%%%%%%%%%%%%%%%%%%%%%%%%%%
\begin{equation}\label{app:eq:TF_Momentum}
  \frac{\partial\,(\varrho_\alpha \uav)}{\partial t}\, +\, \grad\cdot\Big(\varrho_\alpha\uav\uav+\bPia\Big)\, =\,
  \frac{e_\alpha}{m_\alpha}\varrho_\alpha\Big(\Ev+ \frac{\uav}{c}\times\Bv\Big)\,,
\end{equation}
%%%%%%%%%%%%%%%%%%%%%%%%%%%%%%%%%%%%%%%
where $\varrho_\alpha=m_\alpha n_\alpha$, and $m_\alpha$ and $e_\alpha$ are the mass and the electric charge of the species $\alpha$. 
Multiplying (\ref{app:eq:TF_Momentum}) by $e_\alpha/m_\alpha$ and summing over the species index $\alpha$, one obtains
%%%%%%%%%%%%%%%%%%%%%%%%%%%%%%%%%%%%%%%
\begin{equation}\label{app:eq:TF_Ohm-0}
  \frac{\partial}{\partial t}
  \underbrace{\left(\sum_\alpha e_\alpha n_\alpha\uav\right)}_{=\,\Jv}\, +\, \grad\cdot\left(\sum_\alpha\varrho_\alpha\uav\uav
  +\sum_\alpha\frac{e_\alpha}{m_\alpha}\bPia\right)\, =\,
 \sum_\alpha\frac{e_\alpha^2n_\alpha}{m_\alpha}\Big(\Ev+ \frac{\uav}{c}\times\Bv\Big)\,,
\end{equation}
%%%%%%%%%%%%%%%%%%%%%%%%%%%%%%%%%%%%%%%
which, for a quasi-neutral proton-electron plasma $\nee=\nii=n$, rewrites as
%%%%%%%%%%%%%%%%%%%%%%%%%%%%%%%%%%%%%%%
\begin{align}\label{app:eq:TF_Ohm-1}
  \frac{\partial\,\Jv}{\partial t}\, +\,
  e\grad\cdot\Big[n\big(\uiv\uiv-\uev\uev\big)\Big]\,=\, &
  -\frac{e}{\me}\grad\cdot\Big(\bPie-\varepsilon_m\bPii\Big)\,
  +\,\frac{e^2n}{\me}(1+\varepsilon_m)\Ev\nonumber\\
  \,&\,\nonumber\\
  \,&+\,\frac{e^2n}{\me}\left(\frac{\uev+\varepsilon_m\uiv}{c}\times\Bv\right)\,.
\end{align}
%%%%%%%%%%%%%%%%%%%%%%%%%%%%%%%%%%%%%%%
By multiplying (\ref{app:eq:TF_Ohm-1}) by $\me/[(1+\varepsilon_m)e^2n]$ and using the relation $\uev=\uiv-\Jv/en$, one eventually obtains the generalized Ohm's law in (\ref{eq:NHK_Ohm_2}):
%%%%%%%%%%%%%%%%%%%%%%%%%%%%%%%%%%%%%%%
\begin{align}\label{app:eq:TF_Ohm-2}
\Ev\,=\,&
\,-\frac{\uiv}{c}\times\Bv\,
+\,\frac{\Jv\times\Bv}{(1+\varepsilon_m)enc}\,
-\,\frac{\grad\cdot\big(\bPie-\varepsilon_m\bPii\big)}{(1+\varepsilon_m)en}\nonumber\\
\,&\,\nonumber\\
\,&\,+\frac{\varepsilon_m}{1+\varepsilon_m}\frac{\mi}{e^2n}\left[\frac{\partial\Jv}{\partial t}\,
+\,\grad\cdot\left(\Jv\uiv+\uiv\Jv-\frac{\Jv\Jv}{en}\right)\right]\,.
\end{align}
%%%%%%%%%%%%%%%%%%%%%%%%%%%%%%%%%%%%%%%

Now, let us consider the electron momentum equation in the quasi-neutral limit, explicitly solving for $\Ev$ and using electron continuity equation in order to rewrite $\partial_t(\varrho_{\rm e}\uev)+\grad\cdot(\varrho_{\rm e}\uev\uev)=\varrho_{\rm e}[\partial_t\uev+(\uev\cdot\grad)\uev]$:
%%%%%%%%%%%%%%%%%%%%%%%%%%%%%%%%%%%%%%%
\begin{equation}\label{app:eq:TF_Ohm-3}
\Ev\,=\,-\frac{\uev}{c}\times\Bv\,
-\,\frac{\grad\cdot\bPie}{en}\,
-\frac{\me}{e}\left[\frac{\partial\,\uev}{\partial t}+\big(\uev\cdot\grad\big)\uev\right]\,,
\end{equation}
%%%%%%%%%%%%%%%%%%%%%%%%%%%%%%%%%%%%%%%
which is exactly equation (\ref{eq:NHK_Ohm}) of our NHK model.
If we now substitute $\uev=\uiv-\Jv/en$ into the $\uev\times\Bv$ and $\partial_t\uev$ terms, and use the ion momentum equation equation in order to rewrite $\partial_t\uiv$, after some manipulations we then obtain:
%%%%%%%%%%%%%%%%%%%%%%%%%%%%%%%%%%%%%%%
\begin{align}\label{app:eq:TF_Ohm-4}
\Ev\,=\,&
\,-\frac{\uiv}{c}\times\Bv\,
+\,\frac{\Jv\times\Bv}{enc}\,
-\,\frac{\grad\cdot\bPie}{en}\nonumber\\
\,&\,\nonumber\\
\,&\,-\varepsilon_m\left\{\Ev\,
+\,\frac{\uiv}{c}\times\Bv\,
-\,\frac{\grad\cdot\bPii}{en}\,
-\,\frac{\mi}{e^2n}\frac{\partial\,\Jv}{\partial t}\,
-\,\frac{\mi}{en}\grad\cdot\Big[n\big(\uiv\uiv-\uev\uev\big)\Big]\right\}\,,
\end{align}
%%%%%%%%%%%%%%%%%%%%%%%%%%%%%%%%%%%%%%%
which rewrites exactly as equation (\ref{app:eq:TF_Ohm-2}).
This eventually proves the equivalence between using the form (\ref{eq:NHK_Ohm}) or the form (\ref{eq:NHK_Ohm_2}) of the generalized Ohm's law.

\section{Fluid equations: explicit form with indexes}\label{app:FluidEqs_indexes}

Dropping the index ${\rm i}$ for ``ions", Eqs.~(\ref{eq:NHK_continuity})-(\ref{eq:NHK_PressTens}) written by components read:

%%%%%%%%%%%%%%%%%%%%%%%%%%%%%%%%%%%%%%%
\begin{equation}\label{app:eq:NHK_continuity}
\frac{\partial\,\varrho}{\partial t}\, +\, \frac{\partial}{\partial x_i}\big(\varrho u_i\big)\, =\, 0\,,
\end{equation}
%%%%%%%%%%%%%%%%%%%%%%%%%%%%%%%%%%%%%%%
%%%%%%%%%%%%%%%%%%%%%%%%%%%%%%%%%%%%%%%
\begin{equation}\label{app:eq:NHK_IonMomentum}
  \frac{\partial\,(\varrho u_i)}{\partial t}\, +\, \frac{\partial}{\partial x_j}\Big(\varrho u_i u_j+\Pi_{ij}\Big)\, =\,
  en\Big( E_i+ \frac{1}{c}\epsilon_{ijk}u_jB_k\Big)\, +\, n F_i\,,
\end{equation}
%%%%%%%%%%%%%%%%%%%%%%%%%%%%%%%%%%%%%%%
%%%%%%%%%%%%%%%%%%%%%%%%%%%%%%%%%%%%%%%
\begin{equation}\label{app:eq:NHK_PressTens}
  \frac{\partial\,\Pi_{ij}}{\partial t} + \frac{\partial}{\partial x_k}\Big(\Pi_{ij}u_k+Q_{ijk}\Big) 
  + \Pi_{ik}\frac{\partial\,u_j}{\partial x_k} + \Pi_{jk}\frac{\partial\,u_i}{\partial x_k} =
  \Omega_{\rm c}\,\Big(\epsilon_{ilm}\Pi_{jl}b_m+\epsilon_{jlm}\Pi_{il}b_m\Big)\,,
\end{equation}
%%%%%%%%%%%%%%%%%%%%%%%%%%%%%%%%%%%%%%%
where $\epsilon_{ilm}$ is the Levi-Civita symbol and the moments of the distribution function $f$ are defined as usual:
%%%%%%%%%%%%%%%%%%%%%%%%%%%%%%%%%%%%%%%
\begin{equation}\label{app:eq:NHK_rho-def}
  \varrho\,=\,m\, n\,=\, m\int_{-\infty}^{+\infty}f(\xv,\vv)\,{\rm d}^3\vv\,,
\end{equation}
%%%%%%%%%%%%%%%%%%%%%%%%%%%%%%%%%%%%%%%
%%%%%%%%%%%%%%%%%%%%%%%%%%%%%%%%%%%%%%%
\begin{equation}\label{app:eq:NHK_rhou-def}
  \varrho\, u_i\,=\, m\int_{-\infty}^{+\infty}v_i\,f(\xv,\vv)\,{\rm d}^3\vv\,,
\end{equation}
%%%%%%%%%%%%%%%%%%%%%%%%%%%%%%%%%%%%%%%
%%%%%%%%%%%%%%%%%%%%%%%%%%%%%%%%%%%%%%%
\begin{equation}\label{app:eq:NHK_Pi-def}
  \Pi_{ij}\,= m\int_{-\infty}^{+\infty}\,w_i\,w_j\,f(\xv,\vv)\,{\rm d}^3\vv\,,
\end{equation}
%%%%%%%%%%%%%%%%%%%%%%%%%%%%%%%%%%%%%%%
%%%%%%%%%%%%%%%%%%%%%%%%%%%%%%%%%%%%%%%
\begin{equation}\label{app:eq:NHK_Q-def}
  Q_{ijk}\,= m\int_{-\infty}^{+\infty}w_i\,w_j\,w_k\,f(\xv,\vv)\,{\rm d}^3\vv\,,
\end{equation}
%%%%%%%%%%%%%%%%%%%%%%%%%%%%%%%%%%%%%%%
where $\wv=\vv-\uv$ is the random (thermal) component of the particles velocity.

%\bibliography{biblio}


\end{document}